\documentclass[fleqn,usenatbib]{mnras}

\usepackage{newtxtext,newtxmath}

\usepackage[T1]{fontenc}
\usepackage{ae,aecompl}

\usepackage{graphicx}	
\usepackage{amsmath}	
\usepackage{adjustbox}
\usepackage{makecell}

\usepackage{nccmath}
\usepackage{placeins}

\DeclareMathAlphabet{\mathsc}{OT1}{cmr}{m}{sc}
\def\testbx{bx}%
\DeclareRobustCommand{\ion}[2]{%
\relax\ifmmode
\ifx\testbx\f@series
{\mathbf{#1\,\mathsc{#2}}}\else
{\mathrm{#1\,\mathsc{#2}}}\fi
\else\textup{#1\,{\mdseries\textsc{#2}}}%
\fi}

\newcommand{\Oi}{\ion{O}{i}}

\newcommand{\NaI}{\ion{Na}{i}}
\newcommand{\CaII}{\ion{Ca}{ii}}

\newcommand{\Nii}{\ion{N}{ii}}

\newcommand{\SiII}{\ion{Si}{ii}}

\newcommand{\NaiD}{\ion{Na}{i}~\text{D}}

\newcommand{\NaiDTwo}{\ion{Na}{i}~\text{D}_{2}}

\newcommand{\msun}{\mbox{M$_{\odot}$}}

\newcommand{\kms}{\mbox{$\rm{km}\,s^{-1}$}}

\newcommand{\CaIIRatio}{\ion{Ca}{ii}~R_{\textrm{HVF}}}

\title[]{Probing the Progenitors of Type Ia Supernovae using Circumstellar Material Interaction Signatures}

\author[P. Clark et al.]{Peter Clark,$^{1, 2}$
Kate Maguire,$^{3}$
Mattia Bulla,$^{4}$
Llu\'{i}s Galbany,$^{5}$
Mark Sullivan,$^{6}$
\newauthor Joseph P. Anderson$^{7,8}$ and
Stephen J. Smartt$^{2}$
\\
$^{1}$Institute of Cosmology and Gravitation, University of Portsmouth, Portsmouth PO1 3FX, UK. \\
$^{2}$Astrophysics Research Centre, School of Mathematics and Physics, Queen's University Belfast, Belfast BT7 1NN, UK.\\
$^{3}$School of Physics, Trinity College Dublin, Dublin 2, Ireland.\\
$^{4}$The Oskar Klein Centre, Department of Astronomy, Stockholm University, AlbaNova, SE-106 91, Stockholm, Sweden.\\
$^{5}$Institute of Space Sciences (ICE, CSIC), Campus UAB, Carrer de Can Magrans, s/n, E-08193 Barcelona, Spain.\\
$^{6}$School of Physics and Astronomy, University of Southampton, Southampton, SO17 1BJ.\\
$^{7}$European Southern Observatory, Alonso de C\'ordova 3107, Casilla 19, Santiago, Chile\\
$^{8}$Millennium Institute of Astrophysics, Nuncio Monsenor Sotero Sanz 100, Of. 104, Providencia, Santiago, Chile 
}

\date{Accepted 2021 July 13. Received 2021 July 13; in original form 2021 April 30}

\pubyear{2021}

\begin{document}
\label{firstpage}
\pagerange{\pageref{firstpage}--\pageref{lastpage}}
\maketitle

\begin{abstract}
This work aims to study different probes of Type Ia supernova progenitors that have been suggested to be linked to the presence of circumstellar material (CSM). In particular, we have investigated, for the first time, the link between narrow blueshifted \NaiD\ absorption profiles and the presence and strength of the broad high-velocity \CaII\ near infrared triplet absorption features seen in Type Ia supernovae around maximum light. With the probes exploring different distances from the supernova; \NaiD\ > 10$^{17}$cm, high-velocity \CaII\ features < 10$^{15}$cm. For this, we have used a new intermediate-resolution X-shooter spectral sample of 15 Type Ia supernovae. We do not identify a link between these two probes, implying either that, one (or both) is not physically related to the presence of CSM or that the occurrence of CSM at the distance explored by one probe is not linked to its presence at the distance probed by the other. However, the previously identified statistical excess in the presence of blueshifted (over redshifted) \NaiD\ absorption is confirmed in this sample at high significance and is found to be stronger in Type Ia supernovae hosted by late-type galaxies. This excess is difficult to explain as being from an interstellar-medium origin as has been suggested by some recent modelling, as such an origin is not expected to show a bias for blueshifted absorption. However, a circumstellar origin for these features also appears unsatisfactory based on our new results given the lack of link between the two probes of CSM investigated.
\end{abstract}

\begin{keywords}
circumstellar matter -- supernovae:general -- distance scale.
\end{keywords}



\section{Introduction}
\label{Sec:Introduction}

Type Ia supernovae (SNe~Ia) are the explosions of white dwarfs in binary systems, play a major role in our understanding of the expansion rate of the Universe and are essential for constraining the properties of dark energy to high precision (e.g.,~\citealt{scolnic_2018_CompleteLightcurveSample}). Despite this, there are still major gaps in our understanding of their underlying progenitor systems. While the observed rate of SNe~Ia is measurable by sky surveys (e.g.,~\citealt{li_2011_NearbySupernovaRates, frohmaier_2019_VolumetricRateNormal}), the rate of the underlying progenitor channels is much more uncertain, with the likely possibility that there is more than one way to make a SN~Ia (e.g.,~\citealt{maoz_2014_ObservationalCluesProgenitors, jha_2019_ObservationalPropertiesThermonuclear}). Classical SN~Ia models involve an exploding white dwarf that is close to the Chandrasekhar mass, which has accreted matter from a non-degenerate companion star and resulting in a runaway thermonuclear reaction \citep{whelan_1973_BinariesSupernovaeType}. Models involving the explosion of sub-Chandrasekhar mass white dwarfs have also gained popularity and the explosion can occur through a violent or dynamical merger \citep{pakmor_2010_SubluminousTypeIa} or the triggering of a He-shell on the white dwarf surface, which results in a detonation of the core  \citep[`double-detonation';][]{taam_1980_LongtermEvolutionAccreting, livne_1995_ExplosionsSubChandrasekhar, shen_2009_UnstableHeliumShell, fink_2010_DoubledetonationSubChandrasekharSupernovae}. 

One observational method that has been used to distinguish between potential SN~Ia progenitor scenarios is the presence of circumstellar material (CSM), which is thought to be more likely in the case of a non-degenerate companion star \citep{hachisu_1999_NewEvolutionaryPath, hachisu_1999_WideSymbioticChannel}, although may be present in some merger scenarios (e.g.,~\citealt{raskin_2013_TIDALTAILEJECTION}). Time-varying blueshifted \NaiD\ absorption features have been identified in some SNe~Ia \citep{patat_2007_DetectionCircumstellarMaterial, simon_2009_VariableSodiumAbsorption, blondin_2009_SecondCaseVariable, blondin_2017_ErratumSecondCase, stritzinger_2010_DistanceNGC1316, sternberg_2014_MultiepochHighspectralresolutionObservations}, with the interpretation that these were due to outflowing (hence blueshifted) CSM close to the SN (distances of $\sim10^{17}$ cm) and not interstellar material (ISM). Additionally, an excess of blueshifted compared to redshifted $\NaiD$ absorption features has been identified in larger SN~Ia samples, consistent with a CSM origin in $\sim$20 \% of SNe~Ia \citep{sternberg_2011_CircumstellarMaterialType, maguire_2013_StatisticalAnalysisCircumstellar}.

Potential links between the presence of blueshifted $\NaiD$ absorption and the properties of the SNe~Ia themselves (\SiII\ velocity, colour at maximum) have been suggested \citep[e.g.,][]{foley_2012_LINKINGTYPEIa, maguire_2013_StatisticalAnalysisCircumstellar, phillips_2013_SourceDustExtinction}. SNe~Ia displaying blueshifted $\NaiD$ absorption are associated with redder \textit{B-V} colours at maximum and in some cases also have higher \SiII\ velocities (but this trend is not seen in all studies e.g., \citealt{maguire_2013_StatisticalAnalysisCircumstellar}).  A link between SNe~Ia showing high-velocity \SiII\ features and redder \textit{B-V} colours at maximum was also previously suggested by \cite{wang_2009_IMPROVEDDISTANCESTYPE}.  SNe~Ia with stronger \NaiD\ absorption in low-resolution spectra also have a preference for redshifted late-time emission line velocities \citep{forster_2012_EvidenceAsymmetricDistribution}. 

\cite{phillips_2013_SourceDustExtinction} found that SNe~Ia displaying blueshifted \NaiD\ have stronger \NaI\ column densities compared to what would be expected from their colour-derived extinction. This was not the case for their non-blueshifted sample, where the extinction and \NaI\ column densities followed the Milky Way relation. This study concluded that the strong \NaI\ column densities for the blueshifted \NaiD\ sample was potentially evidence for outflowing CSM from these systems but there does not appear to be a direct relation; some SNe~Ia that are identified as having time-varying \NaiD\ (thought to be a clear indicator of CSM interaction) do not follow this relation. 

The ratio of total-to-selective extinction, R$_V$, measured in the Milky Way has a typical value of $\sim$ 3.1 \citep{fitzpatrick_1999_CorrectingEffectsInterstellar}, though a range of values are observed (\citealt{mortsell_2013_CalibratingMilkyWay} found a value $\sim$3 with an uncertainty of around 10\%, whereas  \citealt{nataf_2013_REDDENINGEXTINCTIONGALACTIC} find a R$_V$ value of 2.5 in the direction of the galactic core). 
Much lower values (R$_V$ < 2) have been measured for some highly extincted SNe~Ia \cite[e.g.,][]{amanullah_2014_PECULIAREXTINCTIONLAW, burns_2014_CarnegieSupernovaProject}, with these measurements confirmed through spectropolarimetry \citep{patat_2015_PropertiesExtragalacticDust, zelaya_2017_ContinuumForegroundPolarization}. However, whilst not measuring the parameter in the same way, larger cosmological samples of normal SNe~Ia have identified a range of R$_V$, from R$_V$ $\leq$ 2 \citep[e.g.,][]{astier_2006_SupernovaLegacySurvey, conley_2007_ThereEvidenceHubble}, to values closer to 3 \citep[e.g.,][]{fink_2010_DoubledetonationSubChandrasekharSupernovae, mandel_2011_TypeIaSupernova}. The cause of the low R$_V$ values seen in some SNe~Ia is debated. Some studies have suggested it could be due to multiple scattering in CSM around the SNe~Ia \citep{wang_2005_DustTypeIa, goobar_2008_LowCircumstellarDust,amanullah_2011_PERTURBATIONSSNeIa}, due to enhanced Na abundances in the CSM \citep{phillips_2013_SourceDustExtinction}. Other studies have concluded that it is mainly produced in the ISM \citep{bulla_2018_EstimatingDustDistances, bulla_2018_SheddingLightType}, where cloud-cloud collisions from the radiation pressure generated by the exploding SNe themselves may produce smaller dust grains \citep{hoang_2017_PropertiesAlignmentInterstellar, giang_2020_TimevaryingExtinctionPolarization}.

High-velocity components in the \CaII\ NIR triplet absorption feature are observed in nearly all SNe~Ia spectra obtained around maximum light \citep[][]{hatano_1999_HighVelocityEjecta,gerardy_2004_SN2003duSignatures,mazzali_2005_HighVelocityFeaturesUbiquitous,childress_2013_HighvelocityFeaturesType, childress_2013_SpectroscopicObservationsSn, maguire_2014_ExploringSpectralDiversity}, though the strengths of such components vary significantly between objects. The origin of these high-velocity components (with velocities a few thousand \kms\ greater than the photospheric components) is unclear but it has been proposed that the high-velocity components of the \CaII\ NIR triplet could have contributions from CSM (at distances of 
$\sim$ 10$^{14}$ cm) \citep{gerardy_2004_SN2003duSignatures,mazzali_2005_HighVelocityFeaturesUbiquitous, quimby_2006_SN2005cgExplosion, tanaka_2008_OutermostEjectaType} or they may be intrinsic to the SN ejecta \citep{branch_2006_ComparativeDirectAnalysis, blondin_2012_SPECTROSCOPICDIVERSITYTYPE}. If a clear relation between blueshifted $\NaiD$ absorption and high velocity $\CaII$ absorption were to be established, it would be a strong indication of local CSM interaction during the early phases of SNe~Ia evolution.

The focus of this paper is to investigate the link between two potential probes of the presence of CSM around SNe~Ia; the presence of blueshifted $\NaiD$ absorption features and the properties of \CaII\ high-velocity features. We do this using a new sample of intermediate-resolution spectra of SNe~Ia around maximum combined with the literature samples of \cite{sternberg_2011_CircumstellarMaterialType} and \cite{maguire_2013_StatisticalAnalysisCircumstellar}. Section~\ref{Sec:ObservationsAndData} outlines the observations and reduction techniques employed in the construction of the dataset used in this work, while Section~\ref{Sec:Analysis} details the analysis of the sample of spectra obtained, along with their complementary photometric observations and host galaxy measurements. Section~\ref{Sec:Results} describes the main results and Section~\ref{Sec:Discussion} discusses the implications of these results in the context of the progenitors of SNe~Ia and previous studies. Section~\ref{Sec:Conclusions} presents a summary of our main findings. Throughout this paper, we assume a Hubble constant, H$_0$ of 70 \kms\ Mpc$^{-1}$ and adopt a standard cosmology with $\Omega_M$=0.27 and $\Omega_{\Lambda}$=0.73.

\begin{table*}
\caption{Discovery and host galaxy information for the SN~Ia sample. The discovery survey is that reported on the TNS. Two redshifts are listed, the heliocentric value from NED along with the final adopted redshift. The redshift calibration method for each adopted redshift is listed as either emission features (a mean value if there is more than one feature) seen in the X-shooter spectra or if no lines were seen, the value from NED is used.}
\label{tab:Ia_Host_SummaryTable}
\begin{adjustbox}{max width=\linewidth}
\begin{tabular}{lllllllll}
\hline
SN name    & Discovery  & Host galaxy name           & Host   & NED host $z$ & Adopted $z$ & $z$ calibration                      \\
&survey&&morphology$^a$&&&method\\
\hline
SN~2016hvl & ATLAS$^b$            & UGC~03524             & S$^1$     & 0.0131             & 0.0131  & H$\alpha$                   \\
SN~2016ipf$^{*}$ & ATLAS         & 2MASX J08071352+0540566 & E$^3$     & 0.021         & 0.0206  & H$\alpha$                \\
SN~2017hm  & ASAS-SN$^b$              & MCG-02-30-003         & Sb$^1$    & 0.0213    & 0.0219  & H$\alpha$   \\
SN~2017hn  & POSS$^c$             & UGC~08204             & S0$^1$    & 0.02385       & 0.02385  & NED                    \\
SN~2017yv  & ASAS-SN                 & ESO~375-G018      & S$^3$     & 0.0156    & 0.0158  & H$\alpha$ + $\Nii$       \\
SN~2017awz & ATLAS                   & MCG+04-26-033         & S$^2$     & 0.022     & 0.0221  & H$\alpha$ + $\Nii$      \\
SN~2017azw & ASAS-SN                  & ESO~015-G010          & Dwarf$^3$ & 0.02     & 0.02    & NED                    \\
SN~2017bkc & ASAS-SN             & J17503055-0148023 & S$^2$     & 0.0174        & 0.0163    & H$\alpha$               \\
SN~2017cbv$^{**}$ & DLT40$^d$     & NGC~5643          & Sc$^1$    & 0.003999             & 0.003999& NED                    \\
SN~2017ckq & ATLAS           & ESO~437-G056          & Sbc$^1$   & 0.0099             & 0.0098   & H$\alpha$                        \\
SN~2017ejb & DLT40             & NGC~4696          & E$^{1}$ & 0.00987          & 0.00987 & NED           \\
SN~2017fgc & DLT40                  & NGC~0474              & S0$^1$    & 0.007722    & 0.007722 & NED  \\
SN~2017fzw & DLT40                & NGC~2217          & S0$^1$    & 0.0054        & 0.0054 & NED     \\
SN~2017gah & ATLAS                 & NGC~7187              & S0$^1$    & 0.008906     & 0.008906& NED           \\
SN~2017guh$^{***}$ & ASAS-SN              & ESO~486-G019          & S0$^1$    & 0.015427      & 0.015427 & NED   \\
SN~2017gvp & ASAS-SN          & UGC~12739             & S$^1$     & 0.023             & 0.0229  & H$\alpha$               \\ 
\hline
\end{tabular}
\end{adjustbox}
\begin{flushleft}
$^a$Host galaxy morphology source: $^1$\cite{devaucouleurs_1991_ThirdReferenceCatalogue}, $^2$visually classified as part of this work, $^3$\cite{lauberts_1981_ESOUppsalaSurvey}.\\
$^b$ATLAS: Asteroid Terrestrial-impact Last Alert System \citep{tonry_2018_ATLASHighcadenceAllsky}\\
$^c$ASAS-SN: All-Sky Automated Survey for Supernovae \citep{shappee_2014_AllSkyAutomated}.\\
$^d$POSS: Puckett Observatory Supernova Search 	(http://cometwatch.com/supernovasearch/discoveries.html).\\
$^e$DLT40: Sub-day cadence SN search of galaxies at D<40 Mpc \citep{sand_2018_Overview40Mpc}.\\
$^{*}$No heliocentric host redshift is available on NED, with the host z taken from the original TNS classification report\\
$^{**}$SN~2017cbv is notable for having a significant early excess in the bluer bands of its light curve \citep{hosseinzadeh_2017_EarlyBlueExcess}. It was also investigated by \cite{ferretti_2017_NoEvidenceCircumstellar} exploring if it displayed any time variable \NaiD\ absorption features - none where detected. \\
$^{***}$SN~2017guh is excluded from further analysis due to strong host galaxy contamination.\\
\end{flushleft}
\end{table*}

\section{Observations and Data Reduction}
\label{Sec:ObservationsAndData}

In this section, we describe the new data sample used in this work of maximum light X-shooter spectra of 16 SNe~Ia. In Section \ref{Subsection:Dataset}, we describe the reduction process and telluric correction. In Section \ref{Subsection:lcprop_speclass}, we discuss the estimates of the light curve properties (light curve width, colour at maximum), as well as the spectroscopic sub-classification of each SN~Ia in the sample. In Section \ref{Subsection:Lit_Comparison_Sample}, we discuss additional literature objects that are included to increase the sample size. 

\subsection{X-shooter spectroscopic observations}
\label{Subsection:Dataset}
The primary dataset used in this work consists of maximum light (within approx. one week of peak) medium-resolution X-shooter spectroscopy of 16 SNe~Ia with wavelength coverage of $\sim$3000 to 25000 \AA\ across three arms (UVB, VIS, NIR) \citep{vernet_2011_XshooterNewWide}. The SNe were chosen for observation with X-shooter in an unbiased manner, using the same criteria as \cite{maguire_2013_StatisticalAnalysisCircumstellar} i.e., the target SNe~Ia had already been spectroscopically classified and reported on the Transient Name Server (TNS) prior to maximum light to eliminate the risk of contamination from other types of transient with a redshift < 0.03 to enable a high signal to noise spectrum to be obtained. A summary of the sample, including their discovery information and properties of their host galaxies, is presented in Table \ref{tab:Ia_Host_SummaryTable}. 

The X-shooter spectra were reduced in a standard manner using the ESO Reflex reduction pipeline to produce flux-calibrated spectra \citep{modigliani_2010_XshooterPipeline,freudling_2013_AutomatedDataReduction}. Additional telluric corrections were applied to the VIS and NIR arm of the majority of the sample with the \texttt{MOLECFIT} software \citep{smette_2015_MolecfitGeneralTool,kausch_2015_MolecfitGeneralTool}. A telluric correction was not applied for 2 SNe~Ia (SNe 2017hn and 2017bkc) for the NIR arm because of low signal-to-noise (S/N). SN~2017guh was consistent with the core of its host galaxy and suffered from significant host galaxy contamination and therefore is removed from further analysis. The spectra were corrected for Galactic extinction using the $E(B-V)$ values from \cite{schlafly_2011_MeasuringReddeningSloan} and the extinction law of \cite{fitzpatrick_1999_CorrectingEffectsInterstellar}. The definition of the redshift of each SN is central to this analysis and is discussed further in Section \ref{subsubsec:z_and_velocity_Calibration}.

\begin{table*}
\caption{Summary table of the SN sub-type classification, light curve and spectral properties.}
\label{tab:Spec_and_LC_info}
\begin{adjustbox}{max width = \linewidth}
\begin{tabular}{lccccccccccc}
\hline
SN name     & Sub-type$^a$ & Light curve & $B-V$  & $B-V_{\textrm{spec}}^{b}$         & MJD$^c$      & Spectral & \NaiD\           & \CaII\ H\&K      & $\NaiDTwo$ pEQW  & $\NaiDTwo$ pEQW \\
&& stretch&at peak&&of spectrum&phase~(d)$^{d}$&profile$^e$&profile$^e$&total (\AA)$^{f}$&blue (\AA)$^{g}$\\
\hline
SN~2016hvl & Ia-91T  & 1.14 $\pm$ 0.05 & --            & $-$0.14 $\pm$ 0.23 & 57702.29 & $-$7.80 $\pm$ 0.46 & Blue        & Inconclusive*    & 0.157 $\pm$ 0.023 & 0.143 $\pm$ 0.012 \\
SN~2016ipf & Ia      & --                & --            & $-$0.08 $\pm$ 0.23 & 57725.28 & $-$3 $\pm$ 1       & No Abs.     & Inconclusive*    & --                 & 0.018 $\pm$ 0.018 \\
SN~2017hm  & Ia      & 1.11 $\pm$ 0.03 & --            & 0.11 $\pm$ 0.23  & 57775.33 & 2.94 $\pm$ 0.15    & Blue        & Inconclusive*    & 0.107 $\pm$ 0.022 & 0.083 $\pm$ 0.010 \\
SN~2017hn  & Ia-91T  & 1.44 $\pm$ 0.22 & --            & 0.23 $\pm$ 0.23  & 57773.32 & 2.46 $\pm$ 2.36    & No Abs.     & Inconclusive*    & --                 & --                 \\
SN~2017yv  & Ia      & 1.01 $\pm$ 0.20 & --            & 0.19 $\pm$ 0.23  & 57797.22 & 2.81 $\pm$ 0.72    & Blue        & Blue        & 0.503 $\pm$ 0.005 & 0.400 $\pm$ 0.003 \\
SN~2017awz & Ia-91T  & 0.95 $\pm$ 0.11 & --            & 0.14 $\pm$ 0.23  & 57816.21 & 4.01 $\pm$ 0.60    & Blue \& Red & Blue \& Red & 1.235 $\pm$ 0.014 & 0.797 $\pm$ 0.008 \\
SN~2017azw & Ia      & --                 & --            & $-$0.05 $\pm$ 0.23 & 57816.02 & 1 $\pm$ 1          & No Abs.     & No Abs.     & --                 & --                 \\
SN~2017bkc & Ia      & --                 & --            & 0.23 $\pm$ 0.23  & 57816.36 & 5 $\pm$ 3          & No Abs.     & No Abs.     & 0.066 $\pm$ 0.022 & 0.009 $\pm$ 0.018 \\
SN~2017cbv & Ia      & 1.11 $\pm$ 0.02   & 0.04 $\pm$ 0.02 & $-$0.23 $\pm$ 0.23 & 57837.15 & $-$2.82 $\pm$ 0.04 & Red     & Red        & 0.036 $\pm$ 0.010 & 0.010 $\pm$ 0.004 \\
SN~2017ckq & Ia      & 1.02 $\pm$ 0.03 & 0.00 $\pm$ 0.04 & 0.00 $\pm$ 0.23  & 57853.06 & 2.42 $\pm$ 0.11    & Blue        & Blue        & 0.069 $\pm$ 0.022 & 0.019 $\pm$ 0.011 \\
SN~2017ejb & Ia-91bg & 0.73 $\pm$ 0.03                 & 0.28 $\pm$ 0.06           & 0.45 $\pm$ 0.23  & 57913.05 & $-$2.84 $\pm$ 0.07          & No Abs.     & No Abs.     & --                 & --                 \\
SN~2017fgc & Ia      & 0.92 $\pm$ 0.02 & --            & 0.22 $\pm$ 0.23  & 57961.32 & 2.15 $\pm$ 0.17    & No Abs.     & Blue \& Red & --                 & --                 \\
SN~2017fzw & Ia      & 0.74 $\pm$ 0.01                 & 0.33 $\pm$ 0.06          & 0.16 $\pm$ 0.23  & 57983.39 & $-$3.99 $\pm$ 0.13       & No Abs.     & No Abs.     & --  & --                 \\
SN~2017gah & Ia      & 0.82 $\pm$ 0.01 & 0.13 $\pm$ 0.02 & 0.30 $\pm$ 0.23  & 57982.26 & $-$2.74 $\pm$ 0.08 & No Abs.     & Inconclusive*    & 0.015 $\pm$ 0.018 & 0.022 $\pm$ 0.010 \\
SN~2017gvp & Ia      & 1.19 $\pm$ 0.15 & --            & 0.02 $\pm$ 0.23  & 58026.10 & $-$2.83 $\pm$ 0.75 & Blue \& Red & Blue        & 0.692 $\pm$ 0.018 & 0.374 $\pm$ 0.007\\ \hline
\end{tabular}
\end{adjustbox}
\begin{flushleft}
$^a$Sub-type classification based on SNID \citep{blondin_2007_DeterminingTypeRedshift}.\\
$^{b}$\textit{B-V} colour calculated from X-shooter spectra at that phase.\\
$^{c}$Modified Julian Date of start time of first X-shooter spectral exposure.\\
$^{d}$Relative to \textit{B}-band maximum. Where no light curve is available, the phase was estimated from the best-fit SNID spectral template. \\
$^{e}$The profiles are classified as having components that are `blueshifted', `redshifted', `blue and redshifted' or `no absorption'.\\
$^{f}\NaiDTwo$ pEQW in the range $-$200 -- 200 \kms. Features where the emission is consistent with spectral noise are set to zero. \\
$^{g}\NaiDTwo$ pEQW in the range $-$200 -- 0 \kms. Features where the emission is consistent with spectral noise are set to zero. \\
*No consistent absorption - absorption was seen from only one of the \CaII\ lines.\\
\end{flushleft}
\end{table*}

\subsection{Light curve properties and spectral classification}
\label{Subsection:lcprop_speclass}

Light curves from the Asteroid Terrestrial-impact Last Alert System (ATLAS \citealt{tonry_2018_ATLASHighcadenceAllsky, smith_2020_DesignOperationATLAS}) sky survey were used to constrain the photometric properties of 9 events in the sample (as indicated in Table~\ref{tab:Spec_and_LC_info}). Photometric information for these objects can be retrieved from the ATLAS forced photometry server \citep{shingles_2021_ReleaseATLASForced}. A measure of the light curve width (`stretch') was estimated using the \texttt{SiFTO} light curve fitting code \citep{conley_2008_SiFTOEmpiricalMethod} for these events. The maximum light \textit{B--V} colour was also estimated using \texttt{SiFTO} for two events (SN~2017ckq, SN~2017gah) with multi-band (ATLAS cyan and ATLAS orange) photometry around maximum light.
For three objects in the sample, photometry from the literature was fit using \texttt{SiFTO}: SN~2017cbv \citep{hosseinzadeh_2017_EarlyBlueExcess}, SN~2017ejb and SN~2017fzw (Galbany et al.~in prep.) The `stretch' values, along with available \textit{B--V} colours at maximum, for the sample are presented in Table~\ref{tab:Spec_and_LC_info}. The stretch range ($\sim$0.82--1.44) is consistent with that found for the low-$z$ Palomar Transient Factory (PTF) SN~Ia sample \citep{maguire_2014_ExploringSpectralDiversity}. 

Since light curve colour measurements were only available for five events, \textit{B--V} colours around maximum were also estimated using the X-shooter spectra. These spectra were integrated through \textit{B} and \textit{V} band filters to obtain the colour at the phase of the spectrum. We have excluded SN~2016hvl from our spectral colour comparisons because its spectral phase ($-7.8\pm0.46$~d) is significantly earlier  than the rest of the sample of $-4$ to +5 d from peak. 

The instrument response of X-shooter is relatively stable, with an estimated relative flux uncertainty of five per cent across the spectrum \citep{vernet_2011_XshooterNewWide}. Since we are only interested here in the colour, the absolute flux calibration is not relevant. As a sanity check on the magnitude of this uncertainty, we have compared the photometric \textit{B--V} values obtained at peak against the values measured from X-shooter spectra for four events (SN~2012cg, SN~2012ht, LSQ12dbr, SN~2013aj) from \cite{maguire_2013_StatisticalAnalysisCircumstellar} and find agreement within the uncertainties for the four events. For the five events in our sample with both photometric and spectroscopic colour measurements, we have also compared these values and find that they are in agreement except for a slight discrepancy ($\sim$1-$\sigma$) for SN~2017cbv, which we attribute to the earlier phase of its spectrum. To quantify the additional uncertainty that will be introduced to changes in phase, we have estimated the colour change within a phase range of $\pm4$ d for the well-studied event SN~2017cbv \citep{hosseinzadeh_2017_EarlyBlueExcess} to be $\sim$0.1 mag, which is also consistent with the range for SN~2011fe and SN~2012cg \citep[see Figure~2 of][]{hosseinzadeh_2017_EarlyBlueExcess}. We have added this additional uncertainty due to phase in quadrature (see Table \ref{tab:Spec_and_LC_info}). 

The SNe~Ia in the sample were sub-classified using the SN IDentification \texttt{SNID} code \citep{blondin_2007_DeterminingTypeRedshift} applied to their X-shooter spectra. The bulk of the sample (11/15) were classified as `normal' SNe~Ia, three as 91T-like events, and one as a 91bg-like event (see Table~\ref{tab:Spec_and_LC_info}). SN~2017fzw is classified as a `normal' SN~Ia based on the \texttt{SNID} comparison but has some unusual properties, as will be discussed further in Galbany et al.~(in prep.).

\subsection{Literature comparison sample}
\label{Subsection:Lit_Comparison_Sample}

In order to increase the number of objects available for inclusion in the statistical analysis of this work, the 15 newly presented objects are supplemented by the inclusion of the combined dataset described in \cite{maguire_2013_StatisticalAnalysisCircumstellar}. This additional dataset consists of 16 SNe~Ia observed with X-shooter and 16 SNe~Ia from \cite{sternberg_2011_CircumstellarMaterialType} with similar intermediate-resolution spectra around maximum light. Our combined sample (referred to as the `full' sample) consists of 47 SNe~Ia. We note that the sample of \cite{sternberg_2011_CircumstellarMaterialType} has a weak selection effect, with six of the objects selected based on identification of strong narrow $\NaiD$ absorption features e.g.,~SN~2006X. 

\begin{figure*}
    \centering
    \includegraphics[width=14cm]{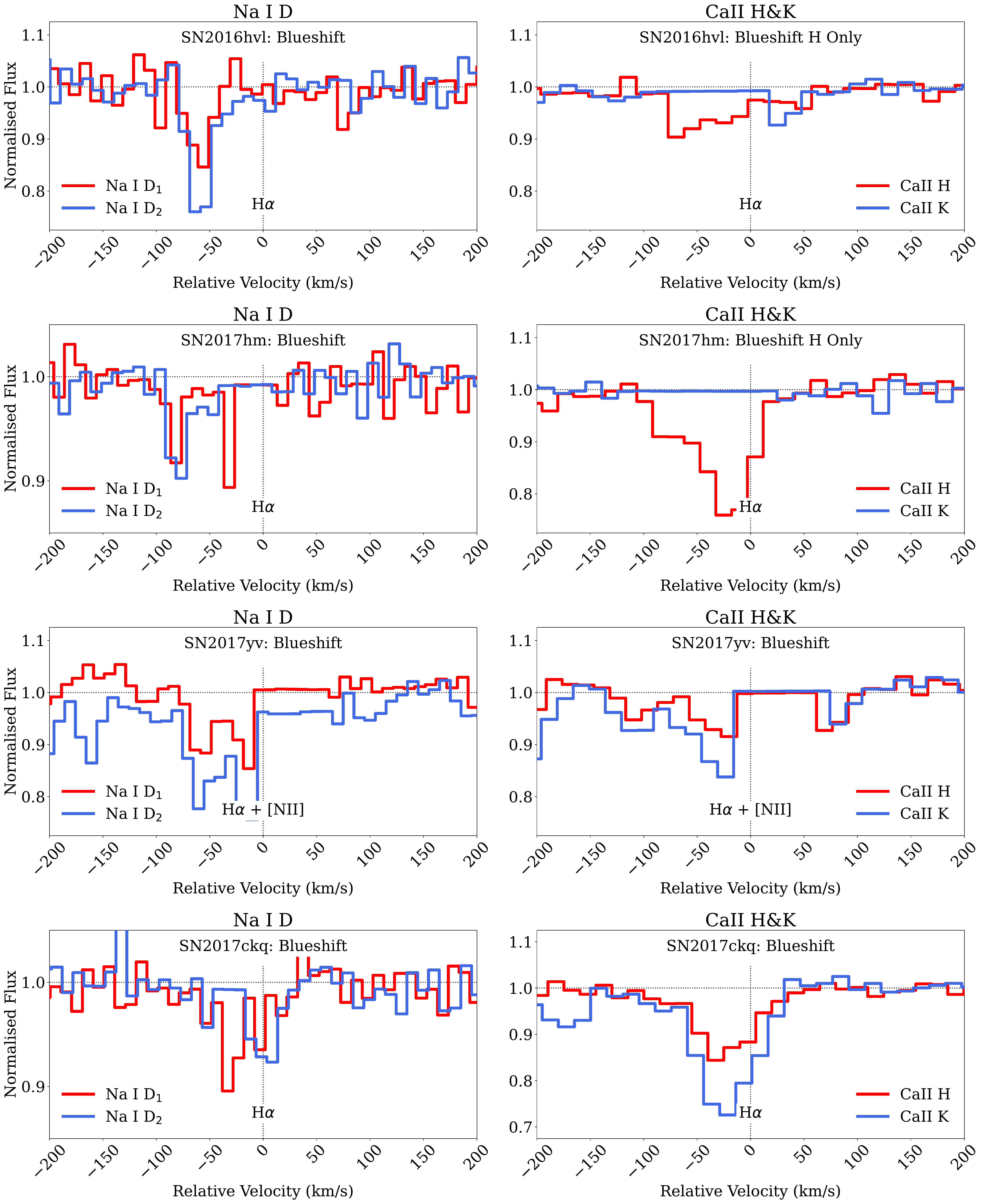}
    \caption{$\NaiD$ (left panels) and \CaII\ H\&K (right panels) regions of objects displaying blueshifted $\NaiD$ absorption features. The method of redshift classification is also indicated below the spectra, with the galactic emission lines used.}
    \label{fig:Ia_Narrow_NaID_Blueshift_Plot}
\end{figure*}

\begin{figure*}
    \centering
    \includegraphics[width=14cm]{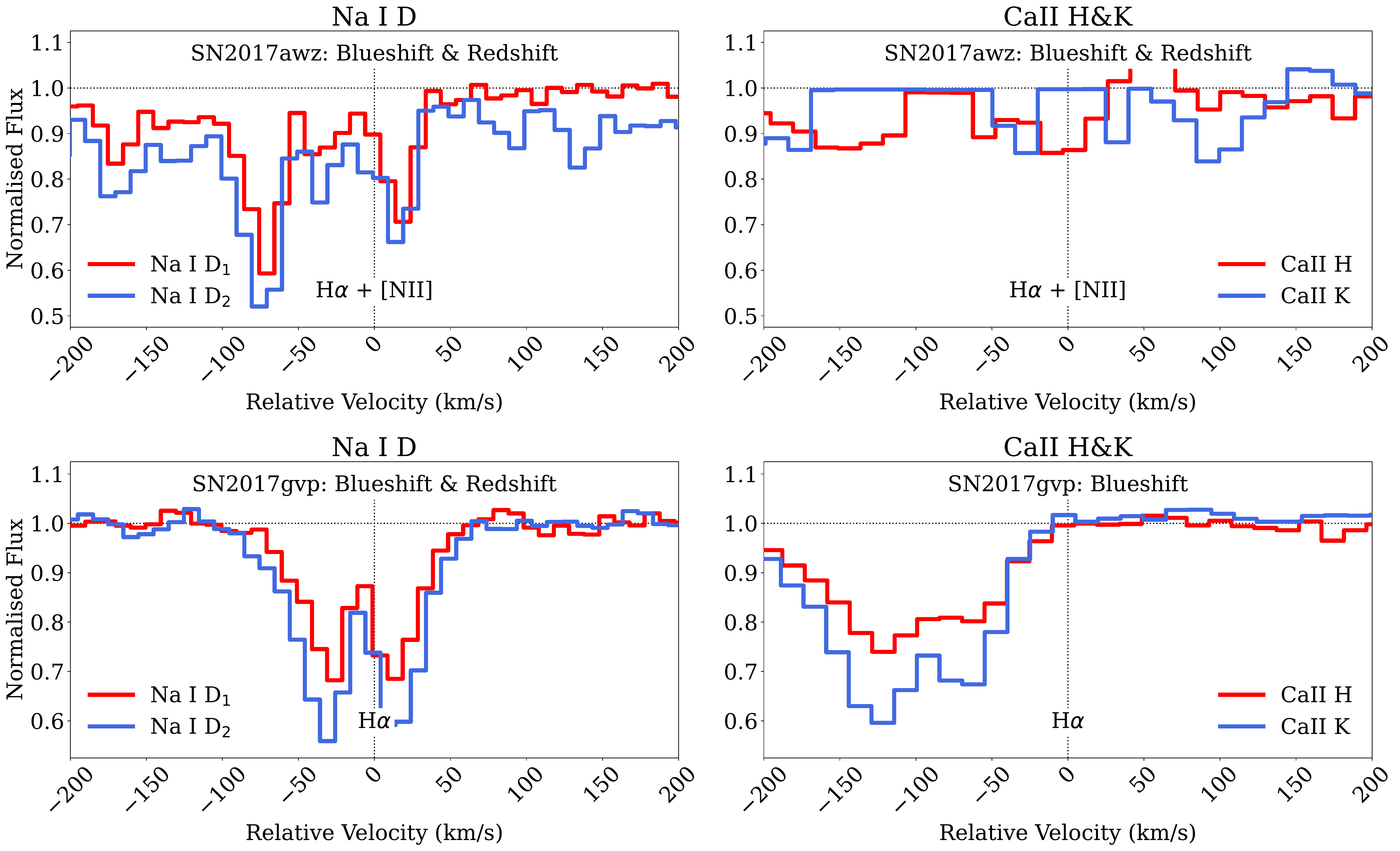}
    \caption{$\NaiD$ (left panel) and \CaII\ H\&K (right panel) regions of objects displaying blue and redshifted $\NaiD$ absorption features. The method of redshift classification is also indicated below the spectra, with the galactic emission lines used.}
    \label{fig:Ia_Narrow_NaID_Blue_and_Redshift_Plot}
\end{figure*}

\begin{figure*}
    \centering
    \includegraphics[width=14cm]{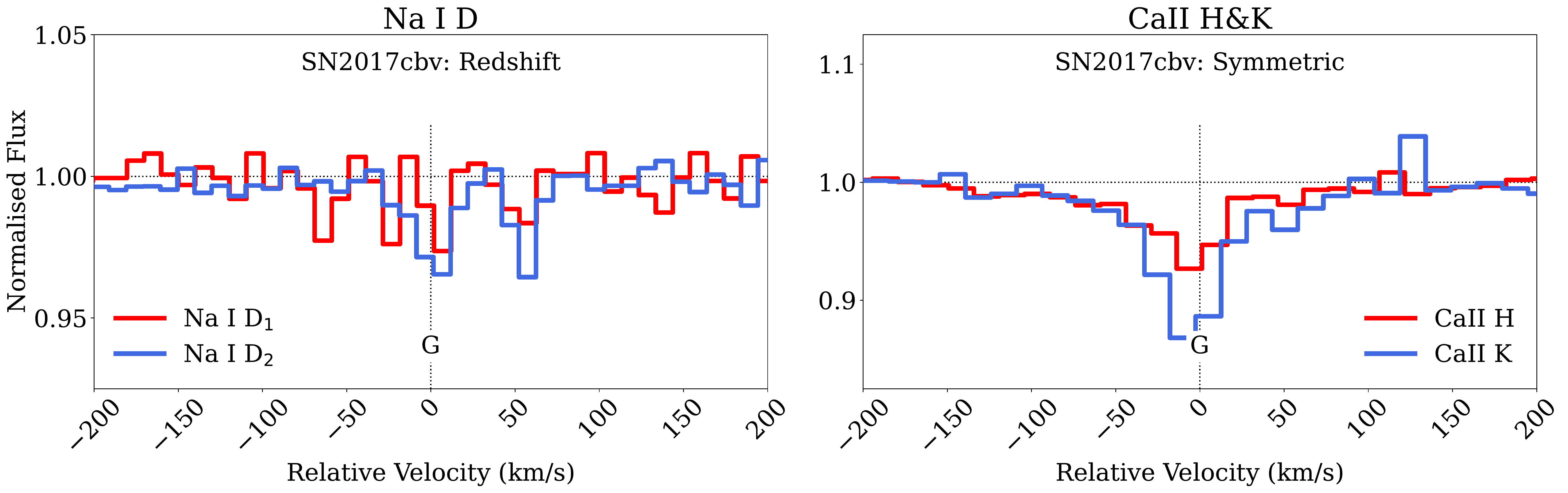}
    \caption{$\NaiD$ (left panel) and \CaII\ H\&K (right panel) regions of SN~2017cbv, the one object displaying redshifted $\NaiD$ absorption features. The method of redshift classification is also indicated below the spectra, with `G' for a galactic redshift estimate from the literature. This classification is supported by the UVES observations described in \citep{ferretti_2017_NoEvidenceCircumstellar}}
    \label{fig:Ia_Narrow_NaID_Redshift_Plot}
\end{figure*}

\begin{figure*}
    \centering
    \includegraphics[width=14cm]{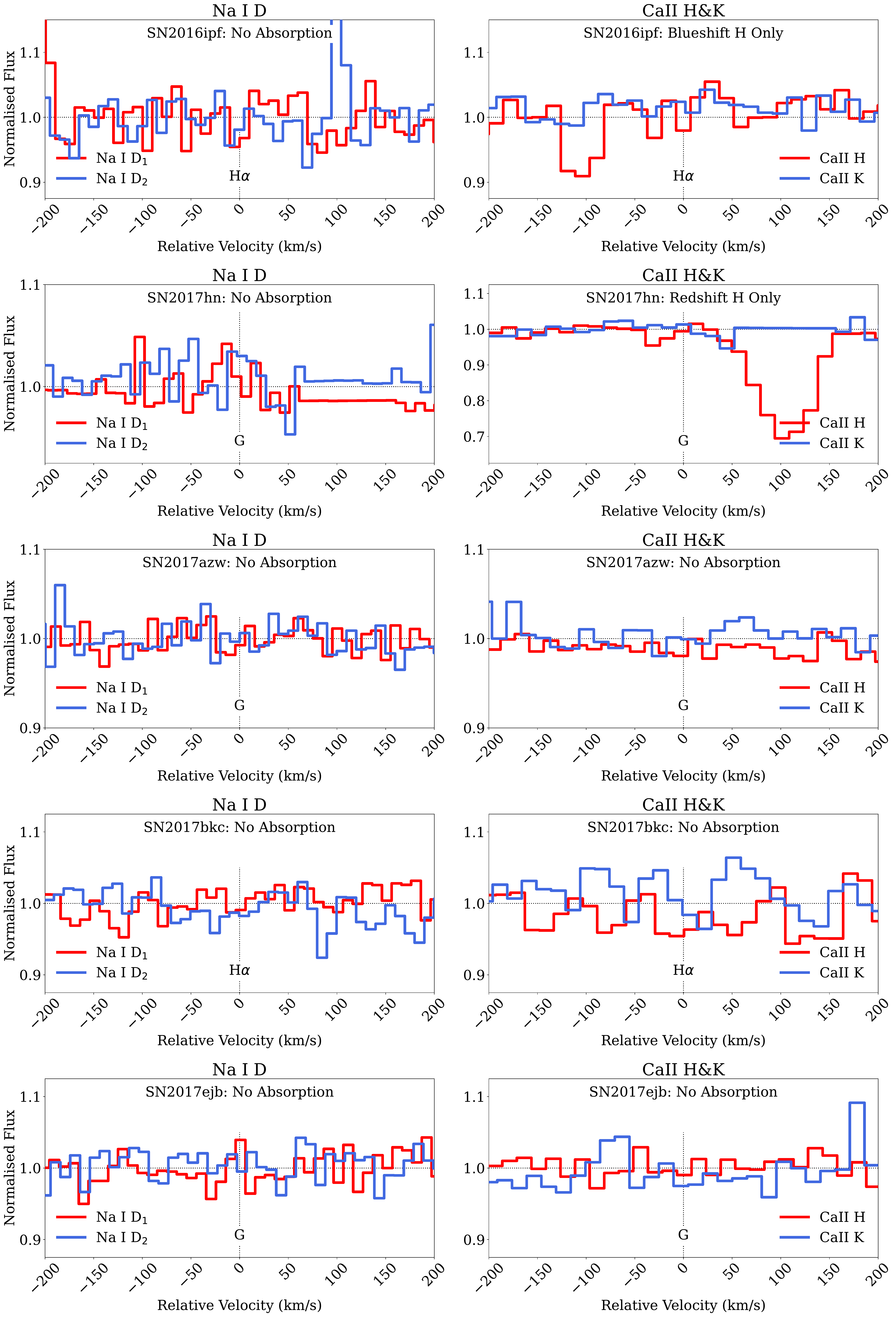}
    \caption{$\NaiD$ (left panel) and \CaII\ H\&K (right panel) regions of objects displaying no $\NaiD$ absorption features. The method of redshift classification is also indicated below the spectra with the galactic emission lines used or `G' for a galactic redshift estimate from the literature.}
    \label{fig:Ia_Narrow_NaID_SNoAbs_Plot_1}
\end{figure*}

\begin{figure*}
    \centering
    \includegraphics[width=14cm]{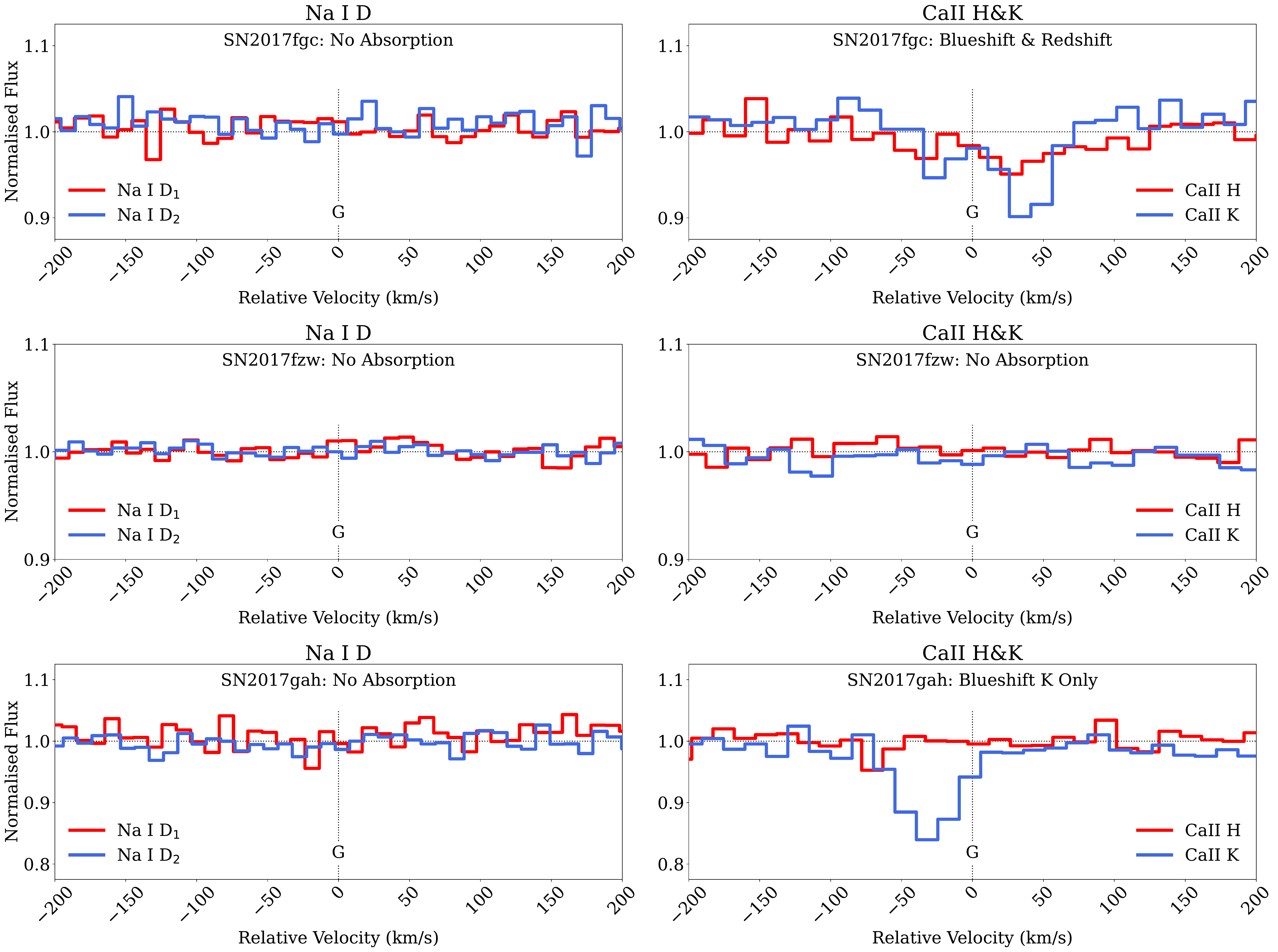}
    \caption{$\NaiD$ (left panels) and \CaII\ H\&K (right panels) regions of additional objects displaying no $\NaiD$ absorption features. The method of redshift classification is also indicated below the spectra with `G' for a galactic redshift estimate from the literature.}
    \label{fig:Ia_Narrow_NaID_SNoAbs_Plot_2}
\end{figure*}

\section{Analysis}
\label{Sec:Analysis}

In this section, we describe the analysis performed to calibrate the spectra to the rest frame and determine the presence of blueshifted material (Section~\ref{Subsection:IdentificationOfBlueShiftedMaterial}), the host galaxy morphologies and stellar mass measurements of the sample (Section~\ref{Subsection:HostGalaxy}), and the measurements of the broad SN absorption components of \SiII\ and \CaII.

\subsection{Identification of narrow absorption features}
\label{Subsection:IdentificationOfBlueShiftedMaterial}

As the narrow features displayed by the $\NaiD$ and \CaII\ H\&K lines are at low velocities, displaying shifts of only a few hundred $\kms$ from their expected rest positions \citep{patat_2007_DetectionCircumstellarMaterial}, accurate redshift calibration is key in determining if any observed features are red or blueshifted. Any contaminating absorption features produced by unrelated interstellar material in the host galaxy would also be expected to be detected within a few hundred $\kms$ of the SN, given the typical velocity distributions of material in galaxies. Such contamination is expected to produce an equal amount of both `blue' and `redshifted' absorption features (with respect to the SN rest-frame) with potential SN-generated features being exclusively blueshifted, due to their origin in out-flowing stellar material and the associated line of sight effects (but see \citealt{hoang_2017_PropertiesAlignmentInterstellar}). 

\subsubsection{Redshift and zero-velocity calibration}
\label{subsubsec:z_and_velocity_Calibration}

To determine the accurate rest frame of each SN, measurements were made using host galaxy emission features identified in the two-dimensional spectral frames, close to the SN position. The host galaxy lines used were the H$\alpha$ and $\Nii$ 6583 \AA\ due to their proximity to the most important feature in this analysis, \NaiD. The lines used in the determination of the redshift for each event and the measured redshift is given in Table \ref{tab:Ia_Host_SummaryTable}. For seven events in the sample, no host galaxy features could be identified in the X-shooter spectra. This was due to either the SN having a significant offset from its host and/or the host being an early-type galaxy that lacks strong emission features. This could potentially result in larger uncertainties due to the lack of an accurate rest frame correction at the SN position. However, for the seven events without identified host features, none showed absorption profiles for \NaiD\ - thus these cases are unaffected. The adopted redshift in Table \ref{tab:Ia_Host_SummaryTable} for these events is the heliocentric redshift retrieved from NED\footnote{The NASA/IPAC Extragalactic Database (NED)
is operated by the Jet Propulsion Laboratory, California Institute of Technology,
under contract with the National Aeronautics and Space Administration.}.  

Based on these redshifts, the `position' of the zero-velocity for each spectrum can be obtained and the classification of the \NaiD\ and \CaII\ H\&K features into `redshifted', `blueshifted', `red and blueshifted' or `no absorption' with respect to the zero rest-frame position from the host galaxy lines can be made. This schema mirrors that used in \cite{maguire_2013_StatisticalAnalysisCircumstellar}, which was modified from that of \cite{sternberg_2011_CircumstellarMaterialType}. We again note that those SNe calibrated using the host redshifts have increased uncertainties due to galactic rotation effects (though no SNe in this sample with observed \NaiD\ absorption was calibrated in this manner). Additionally if the material responsible for producing the narrow galaxy lines is located somewhat in front or behind the true SN position the calibration will also be affected though these offsets are expected to be small with the use of local lines from the 2D spectral frames. Figures~\ref{fig:Ia_Narrow_NaID_Blueshift_Plot} and \ref{fig:Ia_Narrow_NaID_Blue_and_Redshift_Plot} show the SNe with `blueshifted' and `red and blueshifted' \NaiD\ absorption profiles, Figure~\ref{fig:Ia_Narrow_NaID_Redshift_Plot} shows an event (SN~2017cbv) identified with a redshifted \NaiD\ absorption profile (consistent with higher resolution spectra obtained with UVES \citealt{ferretti_2017_NoEvidenceCircumstellar}), and Figs.~\ref{fig:Ia_Narrow_NaID_SNoAbs_Plot_1} and \ref{fig:Ia_Narrow_NaID_SNoAbs_Plot_2} show the objects with `no absorption' components of \NaiD.  
The \CaII\ H\&K features are shown for comparison - in studies of time-varying narrow features, \citep[e.g.,][]{patat_2007_DetectionCircumstellarMaterial} the \CaII\ features are not seen to vary, which is suggested to be due to their higher ionisation potential compared to \NaiD. We find that the \CaII\ classifications are generally in agreement with those of the \NaiD, with just two having different classifications, SN~2017fgc where the \CaII\ H\&K feature had blue and redshifted absorption while the \NaiD\ had no absorption and SN~2017gvp where the \CaII\ H\&K had blueshifted only absorption and the \NaiD\ had blue and red. A similar small number of inconsistencies between these features was also identified in \cite{maguire_2013_StatisticalAnalysisCircumstellar}. As in previous studies, we focus on the \NaiD\ features and only use the \CaII\ H\&K feature for confirming weak \NaiD\ features.

\subsubsection{Continuum calibration and pseudo-equivalent widths}
\label{subsubsec:ContinuumCalibration_pEQWs_IaNarrow}

To make measurements of the pseudo-equivalent widths (pEQWs) of the narrow $\NaiD$ features, any continuum offset needs to be removed. This was done using a linear interpolation between uncontaminated regions on either side of the $\NaiD$ feature. The pEQWs of each feature were then measured following the method of \cite{forster_2012_EvidenceAsymmetricDistribution}. Following previous works, the measurement of the stronger of the two lines making up the \NaiD\ doublet ($\NaiDTwo$) is used as the measure of the feature's overall strength. 

To gain an understanding of overall absorption strength, as well as the strength of only the blueshifted absorption component, the measurement of the pEQW has been made in two velocity regions relative to the calibrated position of zero velocity. The overall strength of the feature measured in the region $-$200 -- +200~\kms\ (`total pEQW') with the pEQW of the blueshifted component measured from $-$200 -- 0~\kms. These values are tabulated in Table \ref{tab:Spec_and_LC_info}. Some previous studies have used larger velocity ranges in the calculation of the pEQWs of \NaiD, such as \cite{forster_2012_EvidenceAsymmetricDistribution} who measured the features out to $\pm$ 300~\kms. However, the higher resolution of the X-shooter spectra allows a narrower region to be used. 

Both these pEQW measurements have been made for the 15 new SNe~Ia explored in this work, including those for which a \NaiD\ absorption feature could not be visually identified, to serve as a confirmation of the non-detection of these features. The regions selected for the continuum were varied randomly within 15~\AA\ of their initially selected locations and the calculation repeated for a total of 5000 measurements to provide a measure of the uncertainty associated with the pEQW measurement. The pEQW values for the literature sample were taken from \cite{maguire_2013_StatisticalAnalysisCircumstellar}, which were obtained using an analogous method.

\subsection{Host galaxy analysis}
\label{Subsection:HostGalaxy}
To search for correlations between the properties of SNe~Ia and their environments, that could in turn be related to differing progenitor scenarios, we perform an analysis of the host properties of our SN sample. This work uses both visual morphological classification of the host galaxies of the SN~Ia sample, along with photometric determinations of the total stellar mass of the hosts, where suitable data are available. We also calculate the projected distance between each SN and the nucleus of its host.

\begin{table}
\centering
\caption{Summary of the host galaxy mass measurements made for the hosts of the combined SNe sample using SDSS photometry.}
\label{tab:HostGalaxyMassMeasurements}
\begin{tabular}{llc}
\hline
SN& Host              & $\log \left(M^* / M_{\odot}\right)$\\ \hline
SN~2006cm&UGC~11723&10.37$\pm$0.07\\
SN~2007af&NGC~5584&9.63$\pm$0.04\\
SN~2007kk&UGC~2828&11.21$\pm$0.01\\
SN~2007le&NGC~7721&10.26$\pm$0.01\\
SN~2008ec&NGC~7469&10.40$\pm$0.03\\
SN~2008hv&NGC~2765&10.70$\pm$0.04\\
SNF~20080514-002&UGC~8472&10.57$\pm$0.05\\
SN~2009ig&NGC~1015&10.27$\pm$0.01\\
SN~2010A&UGC~2019&10.15$\pm$0.01\\
SN~2012cg&NGC~4424&8.83$\pm$0.02\\
SN~2012et&CGCG~476-117&10.34$\pm$0.01\\
SN~2012ht&NGC~3447&8.93$\pm$0.12\\
LSQ12dbr&SDSS~J2058-0258&7.98$\pm$0.15\\
PTF12iiq&(1)&10.57$\pm$0.01\\
SN~2013aj&NGC~5339&9.99$\pm$0.10\\
SN~2013U&(2)&10.34$\pm$0.05\\
SN~2016ipf&(3)&10.22$\pm$0.01\\
SN~2017hn&UGC~08204&10.59$\pm$0.31\\
SN~2017awz&MCG+04-26-033&10.26$\pm$0.01\\
SN~2017fgc&NGC~0474&10.54$\pm$0.02\\
SN~2017gvp&UGC~12739&10.61$\pm$0.09\\ \hline
\end{tabular}
\begin{flushleft}
* Photometric data from \cite{mazzei_2018_GalaxyEvolutionGroups}. \\
(1) 2MASX~J02500784-0016014\\
(2) 2MASX~J10011189+0019458\\
(3) 2MASX~J08071352+0540566\\
\end{flushleft}
\end{table}

\begin{figure}
    \centering
    \includegraphics[width=8.3cm]{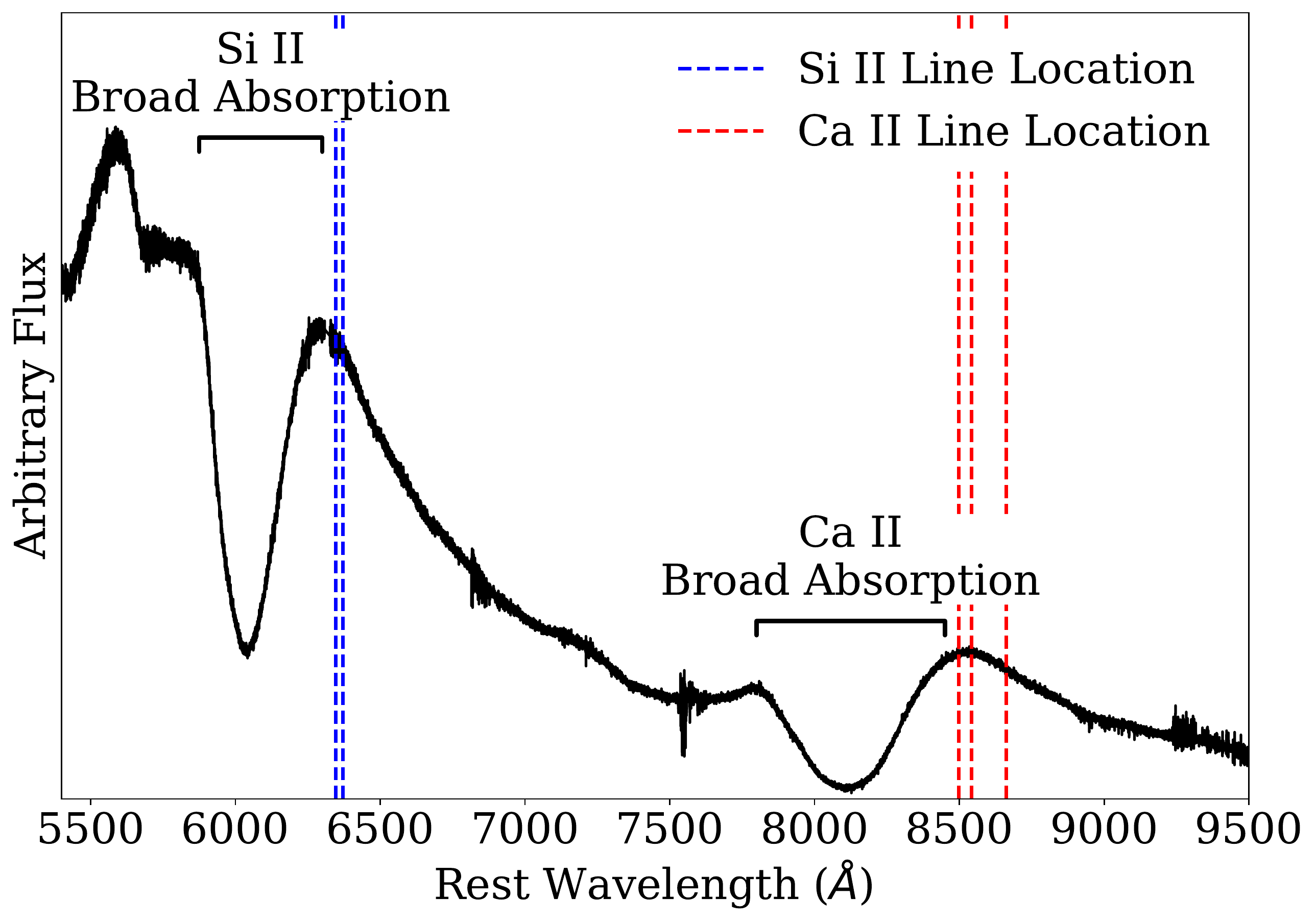}
    \caption{Plot of the 5750--9500 \AA\ region of SN~2017fgc with the \SiII~6355~\AA\ and \CaII\ NIR triplet absorption features identified. The vertical dashed blue lines represent the rest wavelengths of the \SiII\ doublet and the vertical dashed red lines the rest wavelengths of the \CaII\ NIR triplet. }
    \label{fig:Broad_Features_Identified}
\end{figure}

\subsubsection{Host morphological classification}

Host morphological classifications for the SN~Ia sample were retrieved from NED and are sourced from the classification catalogues of \cite{lauberts_1981_ESOUppsalaSurvey} and \cite{devaucouleurs_1991_ThirdReferenceCatalogue}. For two objects (SN~2017awz, SN~2017bkc), visual classification as spiral galaxies was performed as part of this work. These morphological classifications are given in Table~\ref{tab:Ia_Host_SummaryTable} as spiral (`S') and sub-classifications (`S0', `Sa', `Sab', `Sb', `Sbc', `Sc'), elliptical (`E'), or dwarf.

Morphological classifications for the hosts of the SNe included in the combined sample of objects from literature were adopted from their description in \cite{sternberg_2011_CircumstellarMaterialType} or \cite{maguire_2013_StatisticalAnalysisCircumstellar}, apart from four cases, where updated classifications were used. These objects are SN~2008C, which is updated from `S0/a' to `S' based on visual inspection of more recent Pan-STARRS 1 \citep[PS1][]{chambers_2016_PanSTARRS1Surveys} images, SN~2008fp is updated from `Sa' to `S0' \citep{devaucouleurs_1991_ThirdReferenceCatalogue}, SN~2012fw is updated from `S0/a' to `S' based on the classification of \cite{loveday_1996_APMBrightGalaxy}, and PTF12jgb is updated from unclassified to `irregular' based on PS1 imaging.

\subsubsection{Host stellar mass estimates}

To estimate the stellar mass of the hosts of the SNe within this sample, we used photometric measurements of the host galaxies from the Sloan Digital Sky Survey \citep[SDSS;][]{fukugita_1996_SloanDigitalSky, gunn_1998_SloanDigitalSky, gunn_2006_TelescopeSloanDigital, blanton_2017_SloanDigitalSky, doi_2010_PhotometricResponseFunctions} and applied eqn.~8 of \cite{taylor_2011_GalaxyMassAssembly} as was used in \cite{rigault_2018_StrongDependenceType}. This equation provides a method of obtaining the stellar mass based on \textit{g} and \textit{i} band imaging and has an accuracy of $\sim$ 0.1 dex. The most appropriate magnitudes tabulated for SDSS for this analysis were found to be the Petrosian magnitudes \citep{taylor_2011_GalaxyMassAssembly}. Twenty one of the SNe~Ia in our full sample have available photometry in SDSS Data Release 16 \citep{ahumada_2020_16thDataRelease}, and were obtained using the CasJobs data retrieval system. The photometry for the host of SN~2012ht was found to be unreliable due to potential blending with a coincident foreground star, and has been replaced by photometry, also in the SDSS filter system, from \cite{mazzei_2018_GalaxyEvolutionGroups}. The estimated stellar masses using the method of \cite{rigault_2018_StrongDependenceType} are given in Table \ref{tab:HostGalaxyMassMeasurements}. The mean of the log of the host stellar mass of this sample is determined to be 10.3, with a standard deviation of 0.7 dex. The log of the mean host stellar mass from the PTF sample of \cite{pan_2014_HostGalaxiesType} was 10.2 with a standard deviation of 0.8 dex, which is consistent with this sample within the uncertainties. 

\subsubsection{Projected distance from host nucleus}

\begin{figure}
    \centering
    \includegraphics[width=8.25cm]{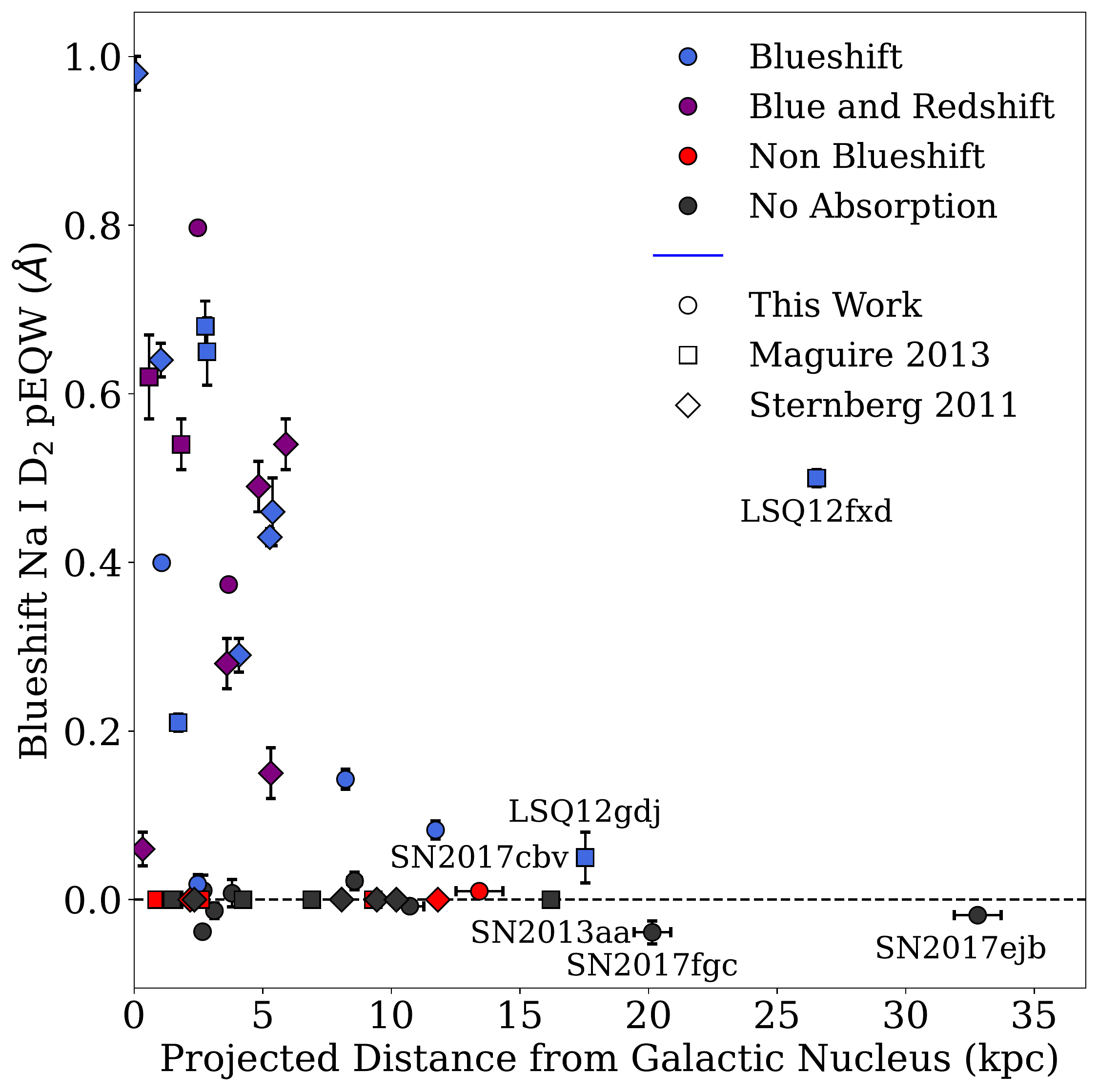}
    \caption{$\NaiDTwo$ pEQW for the full sample vs the measured projected distance between the individual SN and the nucleus of it's host galaxy. SNe with projected distances exceeding 12.5~kpc are labelled.}
    \label{fig:Projected_Distances}
\end{figure}

The projected distance between each SN and the centre of it's host has been determined assuming that SN and host are located at the same distance and are in the same plane i.e., inclination effects are not considered. The full results of these calculations are shown in Figure~\ref{fig:Projected_Distances}.
It is expected that ISM effects will decrease with increasing galactocentric distance making it more likely that objects at large distances showing strong blueshifted \NaiD\ absorption are the result of CSM effects. The majority of the objects within the sample, regardless of \NaiD\ absorption classification are found at projected distances < 12.5 kpc, though two objects (LSQ12fxd and LSQ12gdj) with blueshifted $\NaiD$ absorption are observed at larger distances (26.5 and 17.5 kpc respectively). 
More generally, the means for each \NaiD\ absorption classification are found to be consistent (Blueshift: 6.5$\pm$7.2~kpc, Blue and redshift: 3.2$\pm$1.9~kpc, Non blueshift: 8.9$\pm$8.5~kpc, No absorption: 6.2$\pm$4.6~kpc) indicating no statistical relationship between \NaiD\ absorption classification and the projected distance between SN and host nucleus.

\begin{figure*}
    \centering
    \includegraphics[width=\textwidth]{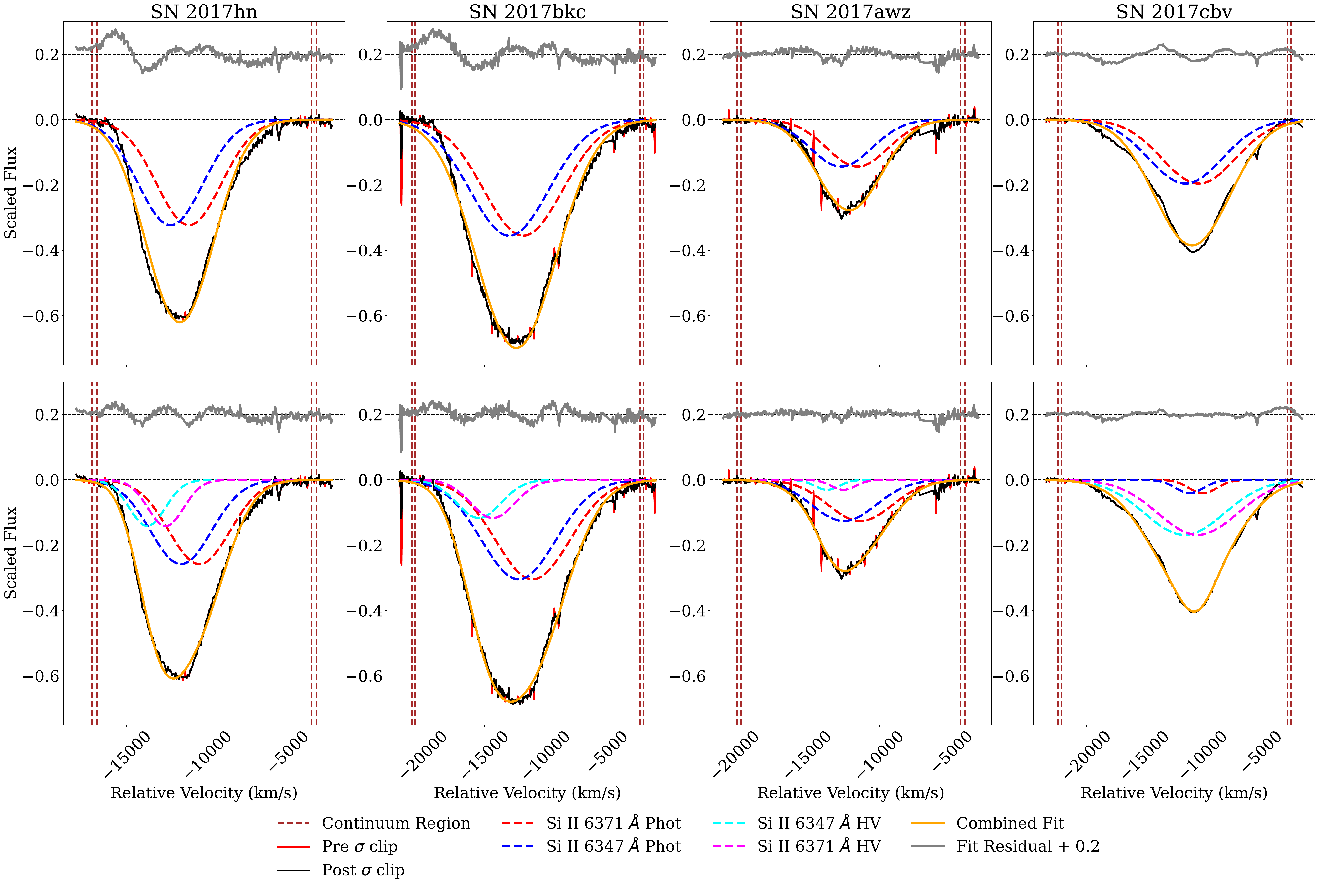}
    \caption{Example \SiII\ fits for a selection of objects. Top Panels: Fitting the feature as being purely photospheric in origin. Bottom Panels: Fitting with the inclusion of a high velocity component. SNe~2017hn and 2017bkc are examples where the fit including a high velocity component reduces the overall residual pattern seen. SN~2017awz is a borderline case where including a high-velocity component provides a small overall contribution to the pEQW of the feature ($\sim$8\%) and was ultimately rejected due to no fitting iteration reaching the required component separation (absorption minima velocities differ by $\sim$ 1200 $\kms$). The fitting of SN~2017cbv is considered to have failed given the very broad nature of the high velocity component, the large high-velocity pEQW contribution ($\sim$88\%) and the small velocity separation between the photospheric and high-velocity components ($\sim$600 $\kms$).}   \label{fig:SiII_Fitting_Examples}
\end{figure*}

\begin{table*}
\caption{Summarised fitting results for the \SiII~6355~\AA\ feature. Means have been weighted using the relevant fitting uncertainties with the uncertainties of these means obtained as the standard deviation of the results. A positive value of $\Delta\bar{\textrm{BIC}}$ indicates a preference for the inclusion of a high velocity feature based on statistical considerations alone.}
\label{tab:SiII_Fitting_Results_New}
\centering
\begin{adjustbox}{max width = \linewidth}
\begin{tabular}{lcccccccc}
\hline
\multicolumn{9}{c}{No High Velocity component} \\ \hline
 &  \multicolumn{3}{c}{Photospheric Component} &  \multicolumn{3}{c}{High Velocity Component} &  &  \\ \hline
SN	&		Min V (\kms)		&		FWHM (\kms)		&		pEQW (\AA)		&		Min V (\kms)		&		FWHM (\kms)		&		pEQW (\AA)		&		$\bar{\chi}^2$		&		$\Delta\bar{\textrm{BIC}}$\\	
\hline
SN~2017awz	&	-11520	$\pm$	50	&	4970	$\pm$	60	&	31.0	$\pm$	0.2	&		-		&		-		&		-		&	0.075	$\pm$	0.003	&		*		\\
SN~2017azw	&	-11820	$\pm$	110	&	6510	$\pm$	70	&	69.6	$\pm$	0.5	&		-		&		-		&		-		&	0.121	$\pm$	0.015	&		*		\\
SN~2017cbv	&	-10370	$\pm$	110	&	7450	$\pm$	50	&	63.5	$\pm$	0.2	&		-		&		-		&		-		&	0.097	$\pm$	0.001	&		*		\\
\hline 
Mean	&	-11320	$\pm$	630	&	6370	$\pm$	1020	&	53.1	$\pm$	16.9	&		-		&		-		&		-		&	0.094	$\pm$	0.019	&	-	\\ \hline

\multicolumn{9}{c}{With High Velocity component} \\ \hline
 &  \multicolumn{3}{c}{Photospheric Component} &  \multicolumn{3}{c}{High Velocity Component} &  &  \\ \hline
SN	&		Min V (\kms)		&		FWHM (\kms)		&		pEQW (\AA)		&		Min V (\kms)		&		FWHM (\kms)		&		pEQW (\AA)		&		$\bar{\chi}^2$		&		$\Delta\bar{\textrm{BIC}}$\\ \hline
SN~2016hvl	&	-7950	$\pm$	200	&	11950	$\pm$	350	&	16.9	$\pm$	0.2	&	-12450	$\pm$	80	&	4100	$\pm$	130	&	2.9	$\pm$	0.2	&	0.100	$\pm$	0.001	&	120	$\pm$	50	\\
SN~2016ipf	&	-9740	$\pm$	100	&	8030	$\pm$	220	&	93.4	$\pm$	3.2	&	-13340	$\pm$	820	&	6350	$\pm$	1270	&	15.8	$\pm$	3.2	&	0.215	$\pm$	0.027	&	100	$\pm$	90	\\
SN~2017hm	&	-9720	$\pm$	260	&	5830	$\pm$	200	&	75.4	$\pm$	7.2	&	-12410	$\pm$	130	&	3280	$\pm$	450	&	17.3	$\pm$	7.2	&	0.052	$\pm$	0.007	&	630	$\pm$	80	\\
SN~2017hn	&	-10470	$\pm$	120	&	4450	$\pm$	60	&	49.6	$\pm$	1.3	&	-12560	$\pm$	120	&	2650	$\pm$	120	&	16.6	$\pm$	1.3	&	0.122	$\pm$	0.006	&	330	$\pm$	50	\\
SN~2017yv	&	-11470	$\pm$	90	&	6930	$\pm$	210	&	97.6	$\pm$	3.1	&	-14290	$\pm$	540	&	3730	$\pm$	1070	&	10.9	$\pm$	3.1	&	0.189	$\pm$	0.073	&	280	$\pm$	200	\\
SN~2017bkc	&	-11110	$\pm$	80	&	6960	$\pm$	100	&	94.6	$\pm$	1.7	&	-14460	$\pm$	100	&	3980	$\pm$	170	&	19.9	$\pm$	1.7	&	0.201	$\pm$	0.016	&	400	$\pm$	30	\\
SN~2017ckq	&	-9880	$\pm$	190	&	5960	$\pm$	120	&	76.2	$\pm$	4.9	&	-12110	$\pm$	70	&	2630	$\pm$	410	&	8.9	$\pm$	4.9	&	0.038	$\pm$	0.006	&	680	$\pm$	100	\\
SN~2017ejb	&	-9070	$\pm$	170	&	7550	$\pm$	60	&	88.1	$\pm$	3.3	&	-11520	$\pm$	110	&	3340	$\pm$	240	&	15.2	$\pm$	3.3	&	0.088	$\pm$	0.030	&	760	$\pm$	100	\\
SN~2017fgc	&	-13940	$\pm$	140	&	9120	$\pm$	100	&	118.6	$\pm$	2.4	&	-18870	$\pm$	120	&	5420	$\pm$	120	&	32.5	$\pm$	2.4	&	0.132	$\pm$	0.004	&	940	$\pm$	40	\\
SN~2017fzw	&	-12190	$\pm$	120	&	8660	$\pm$	90	&	123.4	$\pm$	1.5	&	-17460	$\pm$	110	&	5140	$\pm$	70	&	48.6	$\pm$	1.5	&	0.234	$\pm$	0.008	&	1030	$\pm$	40	\\
SN~2017gah	&	-12100	$\pm$	190	&	10130	$\pm$	260	&	141.6	$\pm$	5.7	&	-15240	$\pm$	820	&	6120	$\pm$	1910	&	18.4	$\pm$	5.7	&	1.744	$\pm$	0.355	&	120	$\pm$	100	\\
SN~2017gvp	&	-10680	$\pm$	50	&	7100	$\pm$	60	&	86.2	$\pm$	0.7	&	-12520	$\pm$	60	&	3420	$\pm$	100	&	9.8	$\pm$	0.7	&	0.082	$\pm$	0.000	&	400	$\pm$	60	\\																																	\hline 												
Mean	&	-10700	$\pm$	1530	&	7080	$\pm$	1950	&	66.7	$\pm$	31.4	&	-13490	$\pm$	2160	&	3770	$\pm$	1290	&	8.7	$\pm$	11.7	&	0.091	$\pm$	0.466	&	520	$\pm$	310	\\ \hline
\end{tabular}
\end{adjustbox}
\begin{flushleft}
* No $\Delta\bar{\textrm{BIC}}$ values are available for these objects as all fitting iterations when including a high-velocity feature were unsuccessful. \\
\end{flushleft}
\end{table*}

\subsection{Line measurements of broad SN features}
\label{subsubsec:BroadFeatureFitting}

One of the main aims of this paper is to determine if there is a relation between  the presence of blueshifted \NaiD\ features and the broad SN features of \SiII~6355~\AA\ and the \CaII\ NIR triplet seen in SNe~Ia around maximum light. The typical identified broad features are shown for SN~2017fgc in Figure~\ref{fig:Broad_Features_Identified}. In particular, we are interested in quantifying the presence of high-velocity components to these features with velocities 1500\kms\ higher than the photospheric contribution, which have been previously attributed to interaction with CSM \citep[e.g.,][]{mazzali_2005_HighVelocityFeaturesUbiquitous}. The identification of a relation would provide evidence of blueshifted \NaiD\ features being intrinsic to the SN system and likely not coming from the ISM.  

\subsubsection{Fitting procedure of the broad \SiII\ and \CaII\ NIR features}

\begin{figure*}
    \centering
    \includegraphics[width=\textwidth]{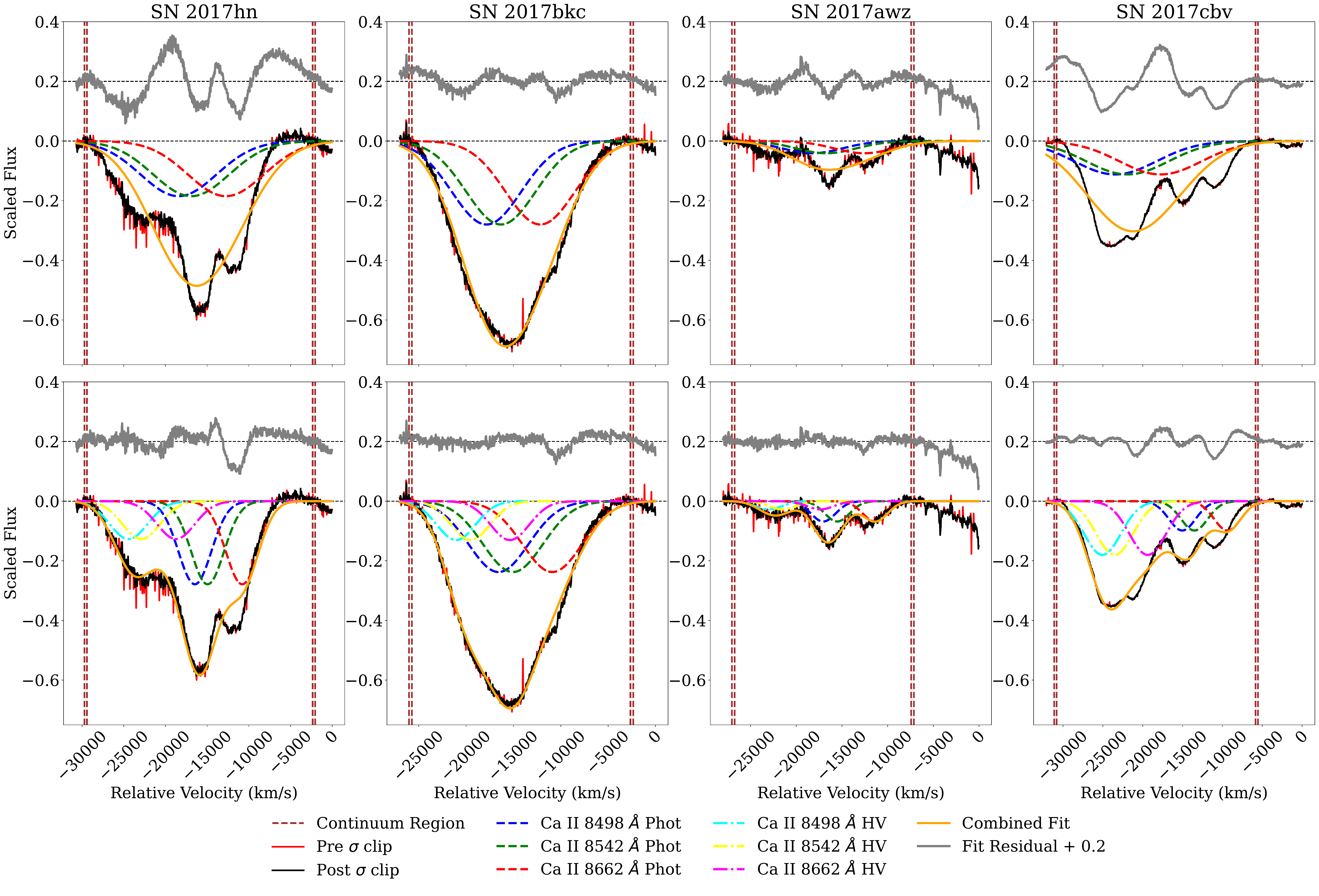}
    \caption{Example \CaII\ fits for a four SNe~Ia. Top panels: Fitting the feature as being purely photospheric in origin. Bottom Panels: Fitting with the inclusion of a high-velocity component. The fitting for all four objects prefer the inclusion of a high-velocity component. However, the features of SNe~2017hn and 2017cbv retain residual patterns indicative of other contributing features e.g., \Oi\ that are not accounted for even with the inclusion of the high velocity \CaII.}
    \label{fig:CaII_Fitting_Examples}
\end{figure*}

\begin{table*}
\caption{Summarised fitting results for the \CaII\ NIR triplet feature. Means have been weighted using the relevant fitting uncertainties with the uncertainties of these means obtained as the standard deviation of the results. A positive value of $\Delta\bar{\textrm{BIC}}$ indicates a preference for the inclusion of a high velocity feature based on statistical considerations alone.}
\label{tab:CaII_Fitting_Results_New}
\centering
\begin{adjustbox}{max width = \linewidth}
\begin{tabular}{lcccccccc}
\hline

 &  \multicolumn{3}{c}{Photospheric Component} &  \multicolumn{3}{c}{High Velocity Component} &  &  \\ \hline
SN	&		Min V (\kms)		&		FWHM (\kms)		&		pEQW (\AA)		&		Min V (\kms)		&		FWHM (\kms)		&		pEQW (\AA)		&		$\bar{\chi}^2$		&		$\Delta\bar{\textrm{BIC}}$		\\\hline
SN~2016ipf	&	-9060	$\pm$	40	&	7620	$\pm$	40	&	86.4	$\pm$	0.2	&	-17890	$\pm$	40	&	5420	$\pm$	40	&	31.8	$\pm$	0.2	&	0.588	$\pm$	0.003	&	560	$\pm$	70	\\
SN~2017hm	&	-9730	$\pm$	40	&	5470	$\pm$	30	&	114.4	$\pm$	0.1	&	-18580	$\pm$	40	&	6140	$\pm$	30	&	87.5	$\pm$	0.1	&	0.609	$\pm$	0.003	&	1500	$\pm$	50	\\
SN~2017hn$^{*}$	&	-10790	$\pm$	110	&	4650	$\pm$	40	&	114.2	$\pm$	0.1	&	-18720	$\pm$	110	&	5300	$\pm$	40	&	58	$\pm$	0.2	&	1.360	$\pm$	0.006	&	1130	$\pm$	50	\\
SN~2017yv	&	-10980	$\pm$	40	&	6350	$\pm$	40	&	126.7	$\pm$	0.1	&	-17590	$\pm$	40	&	5420	$\pm$	40	&	33.6	$\pm$	0.2	&	0.574	$\pm$	0.004	&	490	$\pm$	50	\\
SN~2017awz	&	-11360	$\pm$	40	&	3180	$\pm$	40	&	19.3	$\pm$	0	&	-17460	$\pm$	40	&	3690	$\pm$	40	&	8.9	$\pm$	0.1	&	0.107	$\pm$	0.002	&	710	$\pm$	70	\\
SN~2017azw$^{*}$	&	-11230	$\pm$	110	&	3090	$\pm$	30	&	41.6	$\pm$	0.6	&	-22060	$\pm$	110	&	6300	$\pm$	40	&	97.9	$\pm$	0.1	&	1.582	$\pm$	0.005	&	1120	$\pm$	60	\\
SN~2017bkc	&	-10790	$\pm$	40	&	7380	$\pm$	40	&	152.4	$\pm$	0.6	&	-15380	$\pm$	40	&	5050	$\pm$	40	&	60.3	$\pm$	0.6	&	0.398	$\pm$	0.003	&	310	$\pm$	60	\\
SN~2017cbv$^{*}$	&	-9360	$\pm$	110	&	3960	$\pm$	50	&	34.5	$\pm$	0.3	&	-19370	$\pm$	110	&	5420	$\pm$	40	&	84	$\pm$	0.2	&	0.480	$\pm$	0.003	&	1440	$\pm$	60	\\
SN~2017ckq$^{*}$	&	-9690	$\pm$	40	&	5030	$\pm$	30	&	96.2	$\pm$	0.1	&	-16730	$\pm$	40	&	2810	$\pm$	40	&	19.1	$\pm$	0	&	0.552	$\pm$	0.002	&	830	$\pm$	50	\\
SN~2017ejb$^{*}$	&	-8670	$\pm$	110	&	7180	$\pm$	40	&	137.7	$\pm$	0.4	&	-13450	$\pm$	110	&	4620	$\pm$	30	&	118.4	$\pm$	0.4	&	1.546	$\pm$	0.008	&	580	$\pm$	50	\\
SN~2017fgc	&	-12150	$\pm$	110	&	6340	$\pm$	40	&	102.6	$\pm$	1.1	&	-18670	$\pm$	110	&	7710	$\pm$	50	&	168.4	$\pm$	1.2	&	0.368	$\pm$	0.009	&	850	$\pm$	60	\\
SN~2017fzw	&	-10330	$\pm$	110	&	6570	$\pm$	40	&	123.5	$\pm$	0.6	&	-17750	$\pm$	110	&	7370	$\pm$	40	&	184.3	$\pm$	0.7	&	0.807	$\pm$	0.016	&	1090	$\pm$	60	\\
SN~2017gah$^{*}$	&	-8560	$\pm$	110	&	6090	$\pm$	40	&	117.5	$\pm$	0.3	&	-15790	$\pm$	110	&	7140	$\pm$	40	&	174.4	$\pm$	0.4	&	1.125	$\pm$	0.011	&	900	$\pm$	60	\\
SN~2017gvp	&	-10120	$\pm$	40	&	4970	$\pm$	40	&	78.7	$\pm$	0.1	&	-19860	$\pm$	40	&	6120	$\pm$	40	&	74.7	$\pm$	0.1	&	0.462	$\pm$	0.003	&	1540	$\pm$	50	\\
\\\hline 												
Mean	&	-10210	$\pm$	1040	&	5540	$\pm$	1430	&	76.6	$\pm$	38.6	&	-17730	$\pm$	2020	&	5540	$\pm$	1310	&	57.9	$\pm$	55.5	&	0.621	$\pm$	0.447	&	950	$\pm$	370	\\ \hline	
\end{tabular}
\end{adjustbox}
\begin{flushleft}
$^{*}$ Residual pattern indicative of the presence of an additional feature 
\end{flushleft}
\end{table*}

The \SiII~6355~\AA\ and \CaII\ NIR triplet features were fit using Gaussian profiles broadly following the method of e.g., \cite{childress_2013_SpectroscopicObservationsSn,maguire_2014_ExploringSpectralDiversity}. The spectra were rebinned to 1~\AA\ resolution to increase the S/N and a sigma clipping was applied to remove any residual narrow skylines or other residual artefacts that were not corrected in the reduction process to prevent such features from influencing the fitting. The continuum was removed by selecting regions on either side of the feature and removing with a linear fit as described for the narrow \NaiD\ features. To investigate the presence of high-velocity features in the \SiII~6355~\AA\ and \CaII\ NIR triplet regions, we have fitted both features under two conditions: i) the presence of only photospheric-velocity components and ii) with the additional presence of high-velocity components.

The photospheric components of the \SiII\ feature were fit as a doublet with the relative strengths set assuming the optically thick regime \cite[][]{childress_2013_SpectroscopicObservationsSn,maguire_2014_ExploringSpectralDiversity}, with the full width half-maximum (FWHM) and velocity offset constrained to be the same for both doublet features. The high-velocity components of the \SiII\ were set up in the same manner, with the high-velocity component values independent of the photospheric components.  The same procedure was used for the photospheric and high-velocity components of the \CaII\ NIR triplet, with three linked features for each component.

The weighted mean values of the relevant fit parameters (velocity, FWHM and pEQW) were determined by varying the position of the continuum selection region and refitting 5000 times, with the uncertainties given as the standard deviation of the successful fitting parameter values, combined in quadrature with the redshift and spectral resolution uncertainty for each object. In these fits, the absorption minima velocity of the fitted high-velocity feature must exceed that of the photospheric component by at least 1500~\kms. Based on the results of previous studies, we also make the assumption that the photospheric component is dominant and so the amplitude and width of the high-velocity component must be less than those of the photospheric component. Additionally the widths of both components must exceed 1000~\kms\ to disallow any fits where a narrow noise feature within spectrum has dominated the results. The \CaII\ NIR triplet spectral region is less clean than the \SiII~6355~\AA\ region, with a potential contribution from the \Oi\ 8446~\AA\ feature. We have performed tests including this potential contribution but do not obtain reliable results. Therefore, the \Oi\ 8446~\AA\ feature is not included in our fitting. To ensure consistency of results, the spectra forming the combined dataset of \cite{maguire_2013_StatisticalAnalysisCircumstellar} and \cite{sternberg_2011_CircumstellarMaterialType} have also been fitted using this method where the necessary data are available. These results are presented in Tables \ref{tab:SiII_Fitting_Results_2013Objects} and \ref{tab:CaII_Fitting_Results_2013Objects} for the \SiII\ and \CaII\ NIR features, respectively, for the \cite{maguire_2013_StatisticalAnalysisCircumstellar} sample and in Tables \ref{tab:SiII_Fitting_Results_SternbergObjects} and \ref{tab:CaII_Fitting_Results_SternbergObjects} for \SiII\ and \CaII\ NIR features, respectively, for the \cite{sternberg_2011_CircumstellarMaterialType} sample.

\subsubsection{Determination of the presence of high-velocity components}
\label{highvel_needed}

The presence of a high-velocity \SiII\ feature is determined by the comparison of the fits with and without a high-velocity component using the Bayesian information criterion (BIC, \citealt{schwarz_1978_EstimatingDimensionModel}), where the inclusion of a high-velocity component is strongly favoured if $\Delta\bar{\textrm{BIC}}$ is > 10. Additionally, the fitting of a high-velocity feature must be successful in at least 500 iterations (at least 10 \% of iterations) to prevent the comparisons being unduly weighted by a small number of outlying fits.

The results of the fits for the \SiII\ feature are shown in Table~\ref{tab:SiII_Fitting_Results_New}, with example fits shown in Fig~\ref{fig:SiII_Fitting_Examples}. Three SNe~Ia (SNe~2017awz, 2017azw, 2017cbv) in the sample do not require the inclusion of high-velocity \SiII\ component and are well fit using a photospheric component alone. All remaining objects satisfy the requirements for a high-velocity component based on the BIC and on the number of successful fitting iterations. In a similar manner, the presence of a high-velocity \CaII\ was also determined using the same constraints on BIC and number of successful fitting iterations (Table~\ref{tab:CaII_Fitting_Results_New}). Figure~\ref{fig:CaII_Fitting_Examples} showing example fits to several objects within the sample. A \CaII\ feature was not observed in SN~2016hvl, likely due to its early phase ($-$8~d) and 91T-like classification. All of the remaining 14 objects satisfy the selection criteria and are thus identified as displaying high-velocity \CaII\ absorption. 
\medskip

\begin{figure}
    \includegraphics[width=\columnwidth]{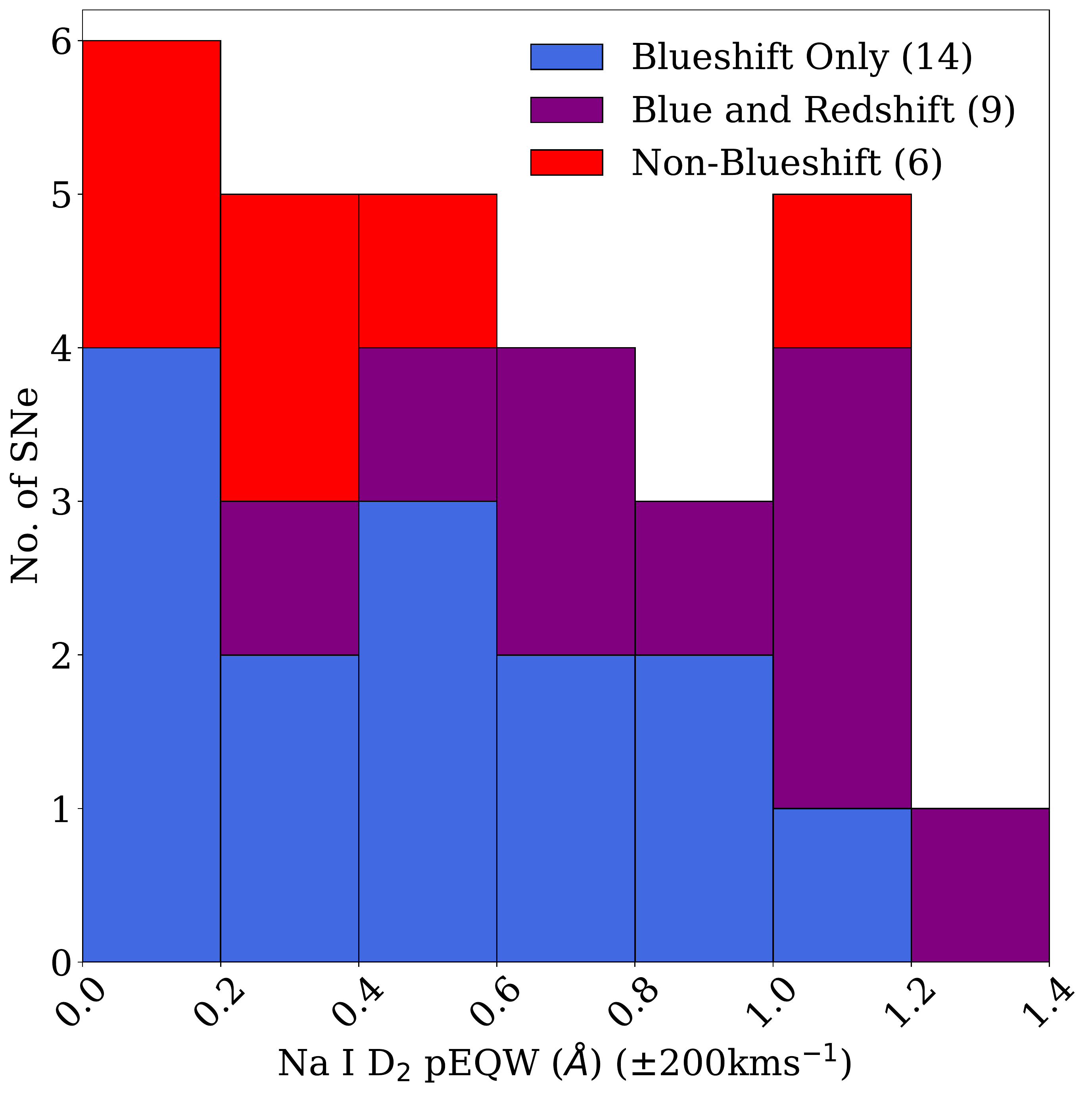}
    \caption{Distribution plots for the measured full feature $\NaiDTwo$ pEQWs. Those objects showing no \NaiD\ absorption features are not included. We find no statistical difference in the pEQW of the blueshifted only and non-blueshift sub-samples.}
    \label{fig:NaID2_PlusMinus200_and_Blue200_Inc2013}
\end{figure}

\section{Results}
\label{Sec:Results}

The following section first outlines the results of the analysis conducted of the broad SN features and the narrow \NaiD\  absorption features. It then describes the connection between the host galaxy properties and the \NaiD\ features, as well as quantifying the observed excess of blueshifted \NaiD\ absorption features. Following this, results of comparisons between the light curve properties and \SiII\ features to the classification of \NaiD\ features are described, before exploring similar comparisons with the properties of the \CaII\ feature.

\begin{figure*}
    \centering
    \includegraphics[width=15cm]{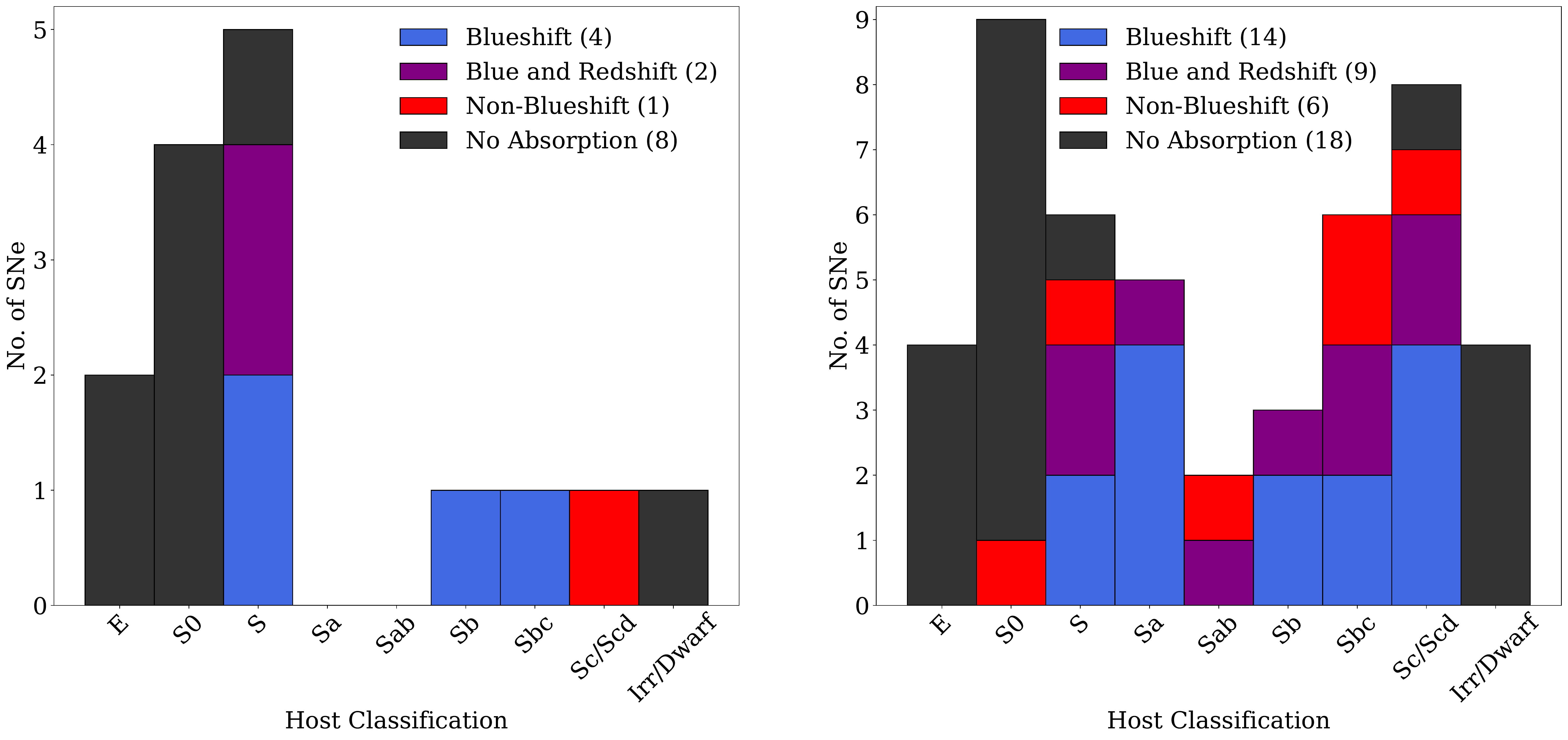}
    \caption{Left panel: Host galaxy morphological distribution of the new SNe~Ia sample discussed in this work. Objects have been divided based on the classification of the $\NaiD$ feature. Right panel: Host galaxy distribution of the combined SNe~Ia sample.}
    \label{fig:Combined_Ia_HostGalaxy_Distribution_Plot}
\end{figure*}

\begin{figure}
    \centering
    \includegraphics[width=8cm]{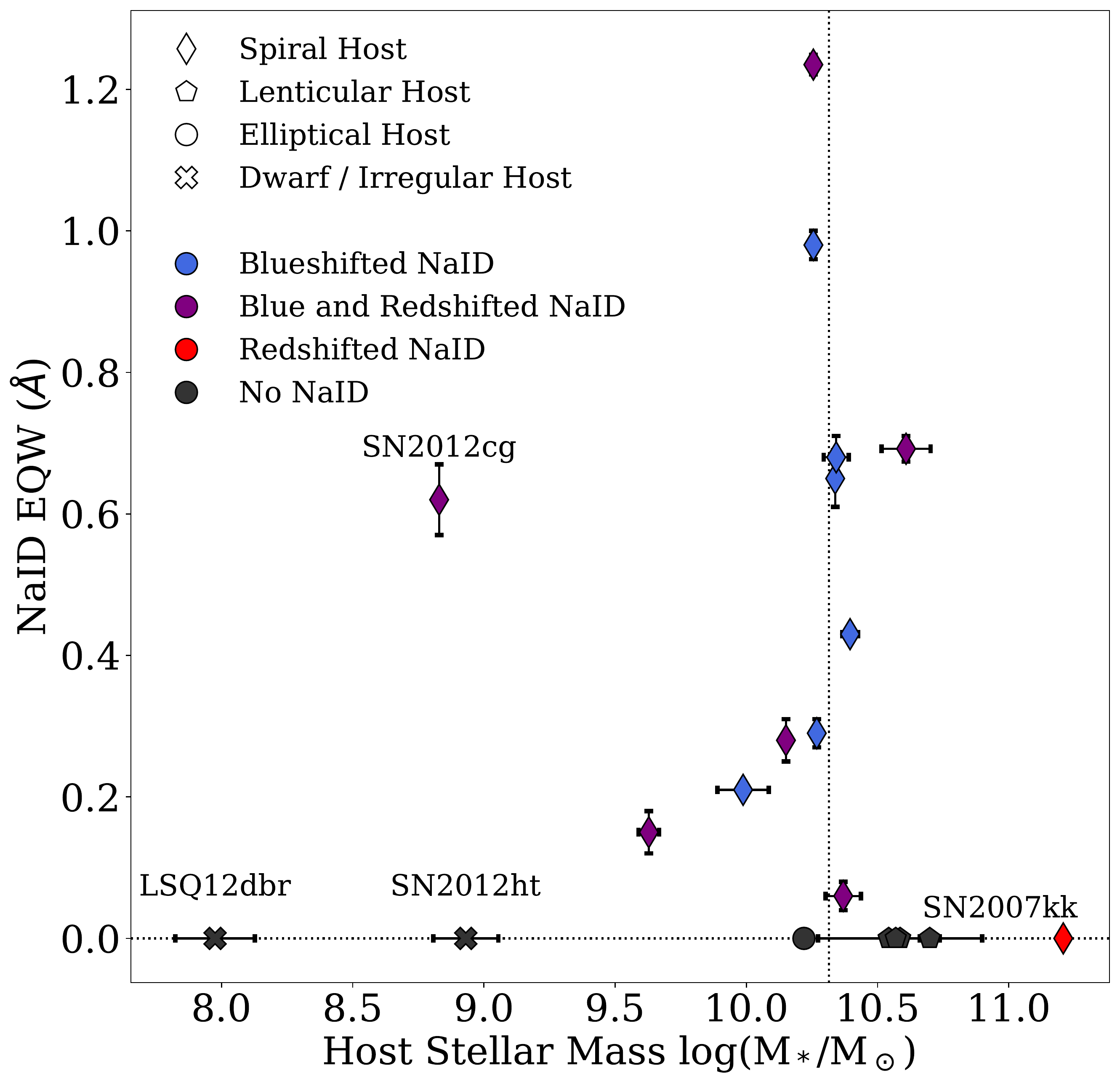}
    \caption{Blueshifted \NaiD\ EQW against host stellar mass. Overall sample mean mass (10.31$\pm$0.71) is indicated by the dashed vertical line. The three lowest mass and the highest mass hosts are labelled. The morphological types are also listed in the legend, along with the \NaiD\ classifications. }
    \label{fig:HostMass_vs_NaIDBlue}
\end{figure}

\subsection{Broad SN features and the presence of high-velocity components}
\label{subsec:BroadFeatureVelocities}

Fitting of the broad \SiII~6635~\AA\ features found that three of the 15 objects (20 \%) in the new sample were well described using a purely photospheric feature, with the rest being better fit with the inclusion of a high-velocity component (Table~\ref{tab:SiII_Fitting_Results_New}). The mean photospheric components (velocity, pEQW, FHWM) of the sub-sample with and without a high-velocity are found to be consistent within the uncertainties.  The fitting of the broad \CaII\ NIR triplet absorption feature was found to heavily favour the inclusion of a high-velocity feature, with all 14 objects for which it was possible to fit the \CaII\ NIR triplet region found to be better described with the inclusion of a high-velocity component. The values measured for the fits are given in Table~\ref{tab:CaII_Fitting_Results_New}. The velocities measured for the \SiII\ and \CaII\ photospheric components are found to be consistent within the uncertainties. However, values given in Table~\ref{tab:CaII_Fitting_Results_New} show that the high-velocity components of the \SiII\ features are narrower and lower velocity than those measured for the \CaII\ NIR triplet.

\subsection{\NaiD\ equivalent widths}
\label{Sec:NaID_EQW}

As outlined in Section~\ref{subsubsec:ContinuumCalibration_pEQWs_IaNarrow}, measurements of the strength for the $\NaiD$ features have been made, using the stronger of the two lines ($\NaiDTwo$) as a proxy for overall strength. Previous studies have suggested that those SNe~Ia showing blueshifted $\NaiD$ absorption have a tendency to display stronger overall $\NaiD$ absorption features (higher pEQW) compared to the rest of the population \citep{phillips_2013_SourceDustExtinction}. 

When those SNe~Ia showing `no absorption' and the `blue and redshifted' absorption are removed from the comparison in the full sample, the remaining objects (i.e. those showing `only blueshifted' or `non blueshifted' \NaiD\ absorption) are well described by a single pEQW distribution. Whilst the median values for the blueshifted and non-blueshifted groups are rather different (0.64 and 0.24 respectively), the corresponding mean values are found to be consistent at 0.46 $\pm$ 0.34 \AA\ and 0.40 $\pm$ 0.39 \AA\ respectively, additionally a K-S test indicates that the two groups are drawn from the same overall population (K-S statistic: 0.333, p-value: 0.657). We note that the non-blueshifted \NaiD\ sub-sample is small with only six objects but we do not identify any clear difference between them. This result is shown in Figure~\ref{fig:NaID2_PlusMinus200_and_Blue200_Inc2013}. As would be expected, this plot also shows that objects showing blue and redshifted \NaiD\ absorption tend to have stronger overall absorption features, which is unsurprising since these objects have multiple absorption components. The pEQWs of the blueshifted components of objects with both blue and redshifted \NaiD\ absorption features are of comparable strength to those of objects showing purely blueshifted \NaiD\ absorption.

\subsection{Host galaxy distribution and connection to narrow \NaiD\ absorption features}
\label{Sec:HostGalaxyDistribution}

Previous studies have shown that SNe~Ia with late-type host galaxies (spiral galaxies) typically have brighter absolute magnitudes and broader light curves (larger `stretch' values), compared to those hosted by early-type galaxies (elliptical and lenticular galaxies), which have been seen to host intrinsically fainter and faster evolving events \citep{hamuy_1995_HubbleDiagramDistant, hamuy_1996_AbsoluteLuminositiesCalan, hamuy_2000_SearchEnvironmentalEffects, riess_1999_ITALBVRIITAL}. It has also been observed that SNe~Ia occurring in late-type galaxies have a higher incidence of blueshifted $\NaiD$ absorption features compared to those in early-type galaxies which are more likely to show no $\NaiD$ absorption features \citep{sternberg_2011_CircumstellarMaterialType,maguire_2013_StatisticalAnalysisCircumstellar}. The sample of new SNe~Ia in this work confirms this previously observed trend with all six of the SNe with blueshifted $\NaiD$ absorption features associated with late-type spiral hosts, compared to just two of the SNe~Ia showing no or non-blueshifted $\NaiD$ absorption features (left panel of Figure~\ref{fig:Combined_Ia_HostGalaxy_Distribution_Plot}). The full combined sample of objects also shows a similar distribution (right panel of Fig~\ref{fig:Combined_Ia_HostGalaxy_Distribution_Plot}).

Figure~\ref{fig:HostMass_vs_NaIDBlue} provides a comparison between the blueshifted $\NaiD$ EQW and the measured host stellar masses of the full sample. Given the reduced number of objects with available host mass estimates, no clear distinctions between the different sub-samples based on $\NaiD$ EQW can be identified, with the log mean host stellar masses of the sub-samples being consistent, 9.93 $\pm$ 0.59 \msun\ (`blue and redshifted'), 10.29 $\pm$ 0.13 \msun\ (`blueshifted only'), 10.36 $\pm$ 0.94 \msun\ (`no absorption'), with a mean of 10.31 $\pm$ 0.72 \msun\ for the full sample.

\subsection{Observed excess of blueshifted \NaiD\ features}
\label{Sec:NaID_BlueExcess}

\begin{table}
\caption{Summary of the \NaiD\ feature classifications of the SN~Ia samples and the corresponding \%EB values.}
\label{Table:NaIDClassification_CombinedSample}
\centering
\begin{adjustbox}{max width = \linewidth}
\begin{tabular}{lcc}
\hline
 & \multicolumn{2}{c}{Number of Events} \\ \hline
Na I D Classification & This Work's Sample & Combined Sample \\ \hline
Blueshifted only & 4 & 14 \\
Redshifted only & 1 & 5 \\
Symmetric & 0 & 1 \\
No Absorption & 8 & 18 \\
 &  &  \\
Total Included & 13 & 38 \\ \hline
\%EB$^{a}$ & 23 & 24 \\ \hline \hline
 &  &  \\
Excluded & & \\ \hline
Blue and Redshifted & 2 & 9 \\
 &  &  \\ \hline
Full Object Total & 15 & 47 \\ \hline
\end{tabular}
\end{adjustbox}
\begin{flushleft}
$^{a}$ The calculation used to determine the percentage excess of blueshifted $\NaiD$ features observed is detailed in Section~\ref{Sec:NaID_BlueExcess}. \\
\end{flushleft}
\end{table}

As previously described, for \NaiD\ features produced by absorbing material within the host galaxy of the SN but not related to the SN itself, it would be expected that an equal number of SNe would display unrelated blueshifted and redshifted \NaiD\ absorption features, as there is no physical reason why material along the same line of sight would have a location bias in front (producing blueshifted features) or behind (producing redshifted features) of a particular SN (but see \citealt{hoang_2017_PropertiesAlignmentInterstellar}). However, previous studies \citep{sternberg_2011_CircumstellarMaterialType, maguire_2013_StatisticalAnalysisCircumstellar} have identified an excess of blueshifted compared to redshifted \NaiD\ components in SN~Ia samples. Using our new sample of 15 SNe~Ia, we have calculated the percentage excess of blueshifted  \NaiD\ features (\%EB) using,

\begin{ceqn}
\begin{align}
    \% E B=\left(\frac{B-R}{B+R+N+S}\right) \times 100
    \label{Eq:ExcessBlueshift}
\end{align}
\end{ceqn}
where B and R are the numbers of events showing exclusively blue and redshifted features respectively, N is the number of events with no observed \NaiD\ absorption features, and S represents the objects displaying a symmetric absorption profile around 0~$\kms$. This is similar to the method of \cite{maguire_2013_StatisticalAnalysisCircumstellar} but for this calculation, any object that displays a combination of red and blueshifted features was excluded given that identifying the source of these features is problematic. \cite{maguire_2013_StatisticalAnalysisCircumstellar} excluded such events from the counts of events showing red or blueshifted features but not from the total number of objects. This exclusion fully removes objects that may have narrow \NaiD\ absorption features from different sources (e.g.,~CSM and host galaxy ISM contamination), which could affect the statistical calculations. 

For the new sample of SNe~Ia in this work, \%EB is found to be 23~\%, while the full sample has an \%EB value of 24~\%. A summary of the classifications of the objects included in the new and full samples is given Table~\ref{Table:NaIDClassification_CombinedSample}. If the probability of a given SN displaying blue- or red-shifted $\NaiD$ features is taken to be equal, as would be expected through purely line-of-sight/host-galaxy contamination, the cumulative binomial probability of observing at least 14 events with blueshifted only $\NaiD$ features compared to five showing purely redshifted $\NaiD$ features is approximately 3.1 per cent.

\begin{table}
\caption{Summary of the \NaiD\ feature classifications of the combined SNe~Ia sample sub-divided by host galaxy classification and the corresponding \%EB values.}
\label{tab:HostGalaxyTypeEB}
\centering
\begin{tabular}{lcc}
\hline
 & \multicolumn{2}{c}{Number of Events} \\ \hline
Na I D Classification & Early-Type Only & Late-Type Only \\ \hline
Blueshifted only & 0 & 14 \\
Redshifted only & 1 & 4 \\
Symmetric & 0 & 1 \\
No Absorption & 12 & 2 \\
 &  &  \\
Total Included & 13 & 21 \\ \hline
\%EB$^{a}$  & -8 & 48 \\ \hline \hline
 &  &  \\
Excluded &  &  \\ \hline
Blue and Redshifted & 0 & 9 \\ 
 &  &  \\ \hline
Full Object Total & 13 & 30 \\ \hline
\end{tabular}
\end{table}

It is expected that SNe in late-type galaxies show $\NaiD$ absorption features more commonly as a consequence of their more gas rich host environments relative to early-type galaxies (with the specifics dependent on the specific local environment of each SN) resulting in additional contamination from unrelated material on coincidental lines of sight to the SN. There is, however, no environmental process that should favour the generation of blue or redshifted features relative to the SN rest frame. To explore this further, \%EB was also calculated separately for those objects with hosts classified as early-type (E and S0) and those with late-type (S), with those hosted by dwarf or irregular galaxies being excluded. Like the previous calculations, objects showing both red and blueshifted features are also excluded. A clear difference between the observed \%EB in early- and late-type galaxies is seen (Table~\ref{tab:HostGalaxyTypeEB}). For early-type galaxies, the negative \%EB ($-$8 per cent) indicates that fewer objects with blueshifted $\NaiD$ features have been observed than would be expected from an even split in distribution. This negative value is likely a result of the small sample size since there are no known processes that favour an overproduction of redshifted narrow $\NaiD$ absorption. Indeed this value is produced by the one detection of a purely redshifted feature in SN~2008fp, further highlighting that early-type galaxies with no absorption features (12 of 13 in this sample) appear to be the norm. In contrast, the \%EB calculated for the SNe~Ia in late-type galaxies shows a strong excess of observed blueshifted features (48 per cent). A corresponding increase in SNe~Ia with redshifted \NaiD\ features is not seen in the late-type host galaxies, as would be expected from isolated ISM contributions alone. The cumulative binomial probability of obtaining this distribution through random chance is approximately 1.5 per cent.

\subsection{Light curve properties and \SiII\ velocities compared to \NaiD\ absorption shifts}
\label{sec:IaStretch}

\begin{figure*}
    \centering
    \includegraphics[width=15cm]{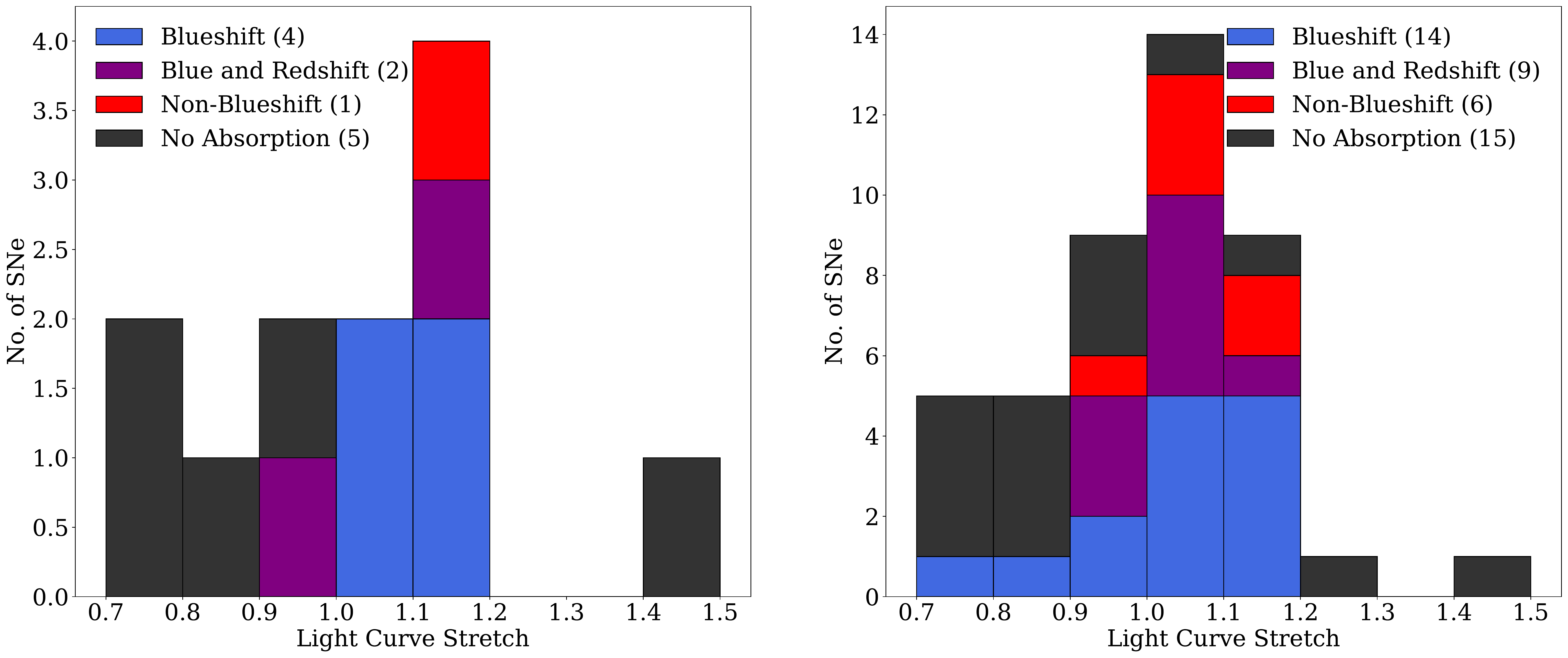}
    \caption{Left panel: Stretch distribution plot of the new SNe~Ia sample discussed in this work. Right panel: Stretch distribution plot of the `full' SNe~Ia sample. The sample is divided based on the properties of their narrow \NaiD\ features.}
    \label{fig:Ia_Stretch_Distribution_Plot}
\end{figure*}

In this section, we investigate the connection between light curve width and \textit{B-V} colour at maximum and the \NaiD\ absorption classifications. The stretch distribution of this new sample alone and the full sample are shown in the left and right panels of  Figure~\ref{fig:Ia_Stretch_Distribution_Plot}, respectively. A K-S test was conducted on the full sample, comparing the measured stretch values of the SNe showing any blueshifted $\NaiD$ component to those showing non-blueshifted or no such features, with the resulting K-S statistic of 0.393 and a p-value of 0.047 indicating that there is a statistically significant difference between the two groups. This is in agreement with the results of \cite{foley_2012_LINKINGTYPEIa} and \cite{maguire_2013_StatisticalAnalysisCircumstellar} where it was observed that SNe~Ia showing no $\NaiD$ absorption features have faster evolving light curves (lower stretch values) than those showing any $\NaiD$ absorption. When the light curve stretch values of the full sample are divided into four corresponding to their $\NaiD$ classification, the weighted mean stretch values are measured to be 1.02 $\pm$ 0.11, 1.03 $\pm$ 0.08, 1.06 $\pm$ 0.05 and 0.86 $\pm$ 0.20 for blueshifted only, blue and redshifted, non-blueshifted, and no $\NaiD$ absorption, respectively. The average light curve stretch value for those SNe~Ia showing no $\NaiD$ absorption features is smaller than the other classifications but they are all marginally consistent within the uncertainties. In agreement with \cite{maguire_2013_StatisticalAnalysisCircumstellar}, we do not identify any relation between the strength (as described by the pEQW) of the blueshifted $\NaiD$ absorption feature of a given SN and its light curve stretch. 

The relationship between light curve width and $\NaiD$ absorption classification is associated with the host distribution of the SNe~Ia within these samples, as those SNe with no $\NaiD$ absorption features largely occur in early-type galaxies, which as discussed in Section~\ref{Sec:HostGalaxyDistribution} are known to host less luminous and faster evolving SNe~Ia. Two of the four SNe~Ia with broad light curves (stretch values exceeding one) and showing no $\NaiD$ absorption (LSQ12dbr and PTF12jgb) occurred within irregular galaxies (that can have histories of recent star formation), with a third (SN~2017hn) hosted by an S0 galaxy, highlighting that the true relationship is more than a simple early/late-type host galaxy split. 

\begin{figure*}
    \centering
    \includegraphics[width=15cm]{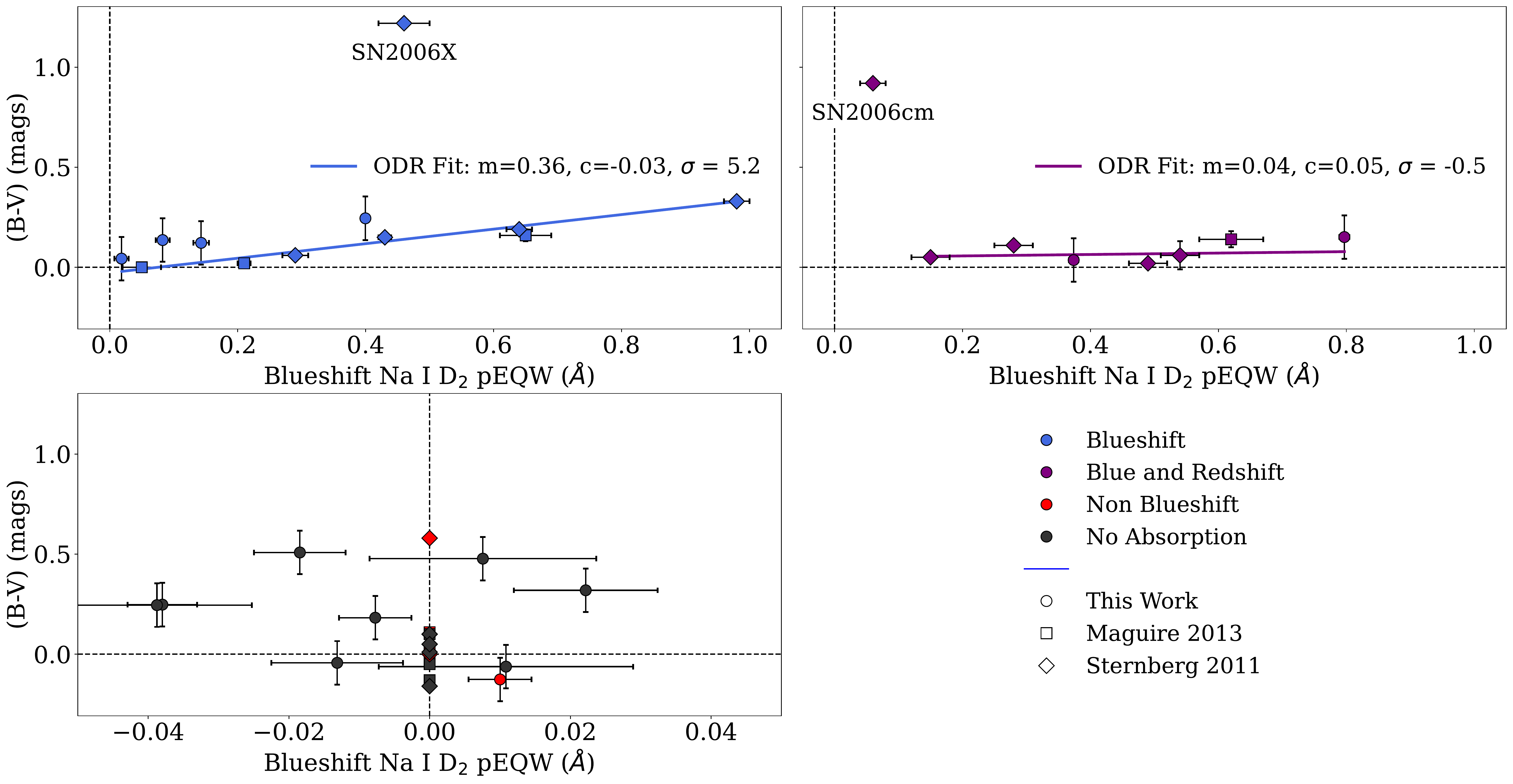}
    \caption{\textit{B}-\textit{V} colour against the blueshifted $\NaiDTwo$ pEQW for the full sample. Top left panel: Objects displaying purely blueshifted \NaiD\ absorption the ODR line of best fit to these points (excluding the extreme outlier SN~2006X) is also shown (solid blue line). Top right panel: Objects displaying  blue and redshifted \NaiD\ absorption the ODR line of best fit to these points excluding the extreme outlier SN~2006cm which like SN~2006X is known to be highly reddened) is also shown (solid purple line) - no trend is identified. Bottom panel: Objects showing no, or non blueshifted \NaiD\ absorption. In all panels the dashed black lines mark the position of zero \textit{B-V} colour and zero blueshifted \NaiD\ absorption.}
    \label{fig:BminusV_Vs_NaID2_200_Blueshift}
\end{figure*}

Previous studies have identified a relation between \textit{B-V} colour at maximum light and the strength of any blueshifted $\NaiD$ absorption features \citep{foley_2012_LINKINGTYPEIa,maguire_2013_StatisticalAnalysisCircumstellar}. A similar investigation of the full sample has been conducted here (Figure~\ref{fig:BminusV_Vs_NaID2_200_Blueshift}) though we fit the those objects showing purely blueshifted \NaiD\ absorption separately from those showing both blue and redshifted \NaiD\ absorption. This analysis reveals the previously found result that SNe~Ia with stronger blueshifted $\NaiD$ absorption features (higher pEQWs) display larger \textit{B}-\textit{V} values (i.e.,~redder colours) is present only for objects showing purely blushifted \NaiD\ absorption with no such trend observed for objects displaying blue and redshifted absorption. The fitted relation between \textit{B-V} colour at maximum and the pEQW of the blueshifted $\NaiDTwo$ \NaiD\ for those objects classified as showing `blueshifted' \NaiD\ absorption retrieves the correlation at 5.2-$\sigma$ significance (provided by the ODR fitting routine). We discuss the possible origin of this correlation further in Section~\ref{sec:CSM_vs_ISM}.

We also explored if there was a correlation between the measured velocity of the \SiII\ features and \NaiD\ pEQW. The full sample confirms the result of \cite{maguire_2013_StatisticalAnalysisCircumstellar} that no such correlation is observed for either the photospheric or high velocity (where present) \SiII\ features.

\subsection{\CaII\ absorption properties compared to \NaiD\ absorption, light curve `stretch' and SN colour}
\label{sec:Broad_Narrow_Comparison}

\begin{figure*}
    \centering
    \includegraphics[width=16cm]{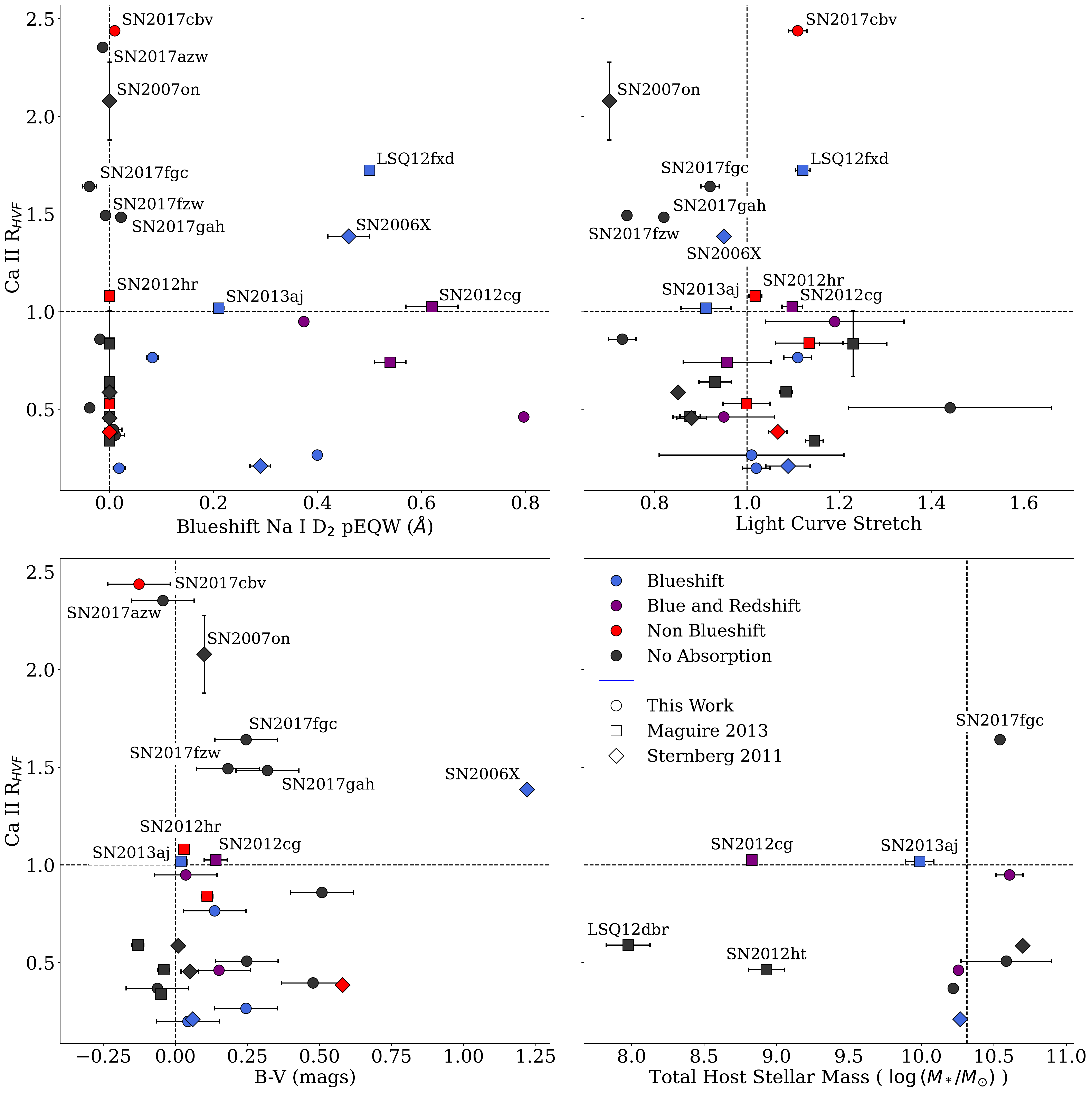}
    \caption{In all panels, the dashed horizontal line indicates the divide between those objects showing strong HV \CaII\ (>1) and the remainder of the sample. The sample is broken down into the original source and type of \NaiD\ shift as shown in the legend. Upper left panel: Comparison between the $\CaIIRatio$ ratio and $\NaiDTwo$ pEQW measurements for our full sample. The dashed vertical line at \NaiD\ absorption strength of zero is present to guide the eye. Upper right panel: Comparison between the $\CaIIRatio$ ratio and light curve stretch. The dashed vertical line at stretch = 1 is present to guide the eye. Lower left panel: Comparison between the $\CaIIRatio$ ratio and SN \textit{B-V} colour at maximum. The dashed vertical line at \textit{B-V} = 0 loosely divides the sample into blue and red objects. Lower right panel: Comparison between the $\CaIIRatio$ ratio and the measured total host galaxy mass of each object where available. The dashed vertical line in this panel marks the mean mass of the sample.}
    \label{fig:CaII_Ratio_Plot_Inc2013}
\end{figure*}

Previous studies \citep[e.g.,][]{mazzali_2005_HighVelocityFeaturesUbiquitous,childress_2013_HighvelocityFeaturesType, maguire_2014_ExploringSpectralDiversity}, have shown that high-velocity components of the \CaII\ NIR absorption features of SNe~Ia obtained near the time of maximum light are very common. There are previous suggestions that they could be due to CSM \citep{gerardy_2004_SN2003duSignatures,mazzali_2005_HighVelocityFeaturesUbiquitous, quimby_2006_SN2005cgExplosion, tanaka_2008_OutermostEjectaType} or may be intrinsic to the SN itself \citep{branch_2006_ComparativeDirectAnalysis, blondin_2012_SPECTROSCOPICDIVERSITYTYPE}. Here we have found that 14 of the 15 objects for which we could obtain measurements display such features (93 per cent). The non-detection in one event (SN~2016hvl) is likely due to a combination of its early phase ($\sim$ 8 d prior to maximum light) along with its classification as a 91T-like event, which typically show weaker absorption features from intermediate-mass elements. As the high-velocity \CaII\ absorption features are so common, it is impossible to link the simple occurrence of such features to the properties of any $\NaiD$ absorption features, blueshifted or otherwise. Therefore, to examine any potential link between the strength of blueshifted $\NaiD$ absorption and high velocity \CaII\ features, the detailed properties of the \CaII\ features must be examined.

Following the method of \cite{childress_2013_HighvelocityFeaturesType}, the strength of the \CaII\ high velocity features is characterised though the ratio of the pEQWs of the high- to photospheric-velocity components ($\CaIIRatio$), with objects having a ratio exceeding one defined as those showing `strong' high velocity \CaII. In the interpretation that high-velocity \CaII\ features come from the same CSM that may be responsible for the blueshifted narrow \NaiD\ features, then a connection between the $\CaIIRatio$ around peak and the pEQW of the blueshifted narrow \NaiD\ features should be seen, with SNe~Ia with the strongest high velocity \CaII\ features also showing the highest blueshifted $\NaiDTwo$ pEQWs. We show this comparison for our full sample in the upper left panel of Figure~\ref{fig:CaII_Ratio_Plot_Inc2013} but identify no trend between the strength of narrow blueshifted $\NaiD$ absorption and the strength of the $\CaIIRatio$, with those SNe~Ia classified as showing blueshifted $\NaiDTwo$ absorption features having a wide range of $\CaIIRatio$. There are also a number of SNe~Ia in the sample with no \NaiD\ absorption that display $\CaIIRatio$ values about one, which is difficult to explain in scenario where both of these probes suggest interaction with CSM.

We have also investigated any potential link between the $\CaIIRatio$ and the light curve stretch, \textit{B-V} colour at maximum, and the stellar mass of the host galaxy for the sample.  Previous studies \citep{childress_2013_HighvelocityFeaturesType, maguire_2013_StatisticalAnalysisCircumstellar} identified a relation between light curve width and $\CaIIRatio$ ratio, with SNe~Ia with high $\CaIIRatio$ ratios also having broader light curves. We do not see this trend for our sample, with a significant fraction of SNe~Ia with high $\CaIIRatio$ ratios having narrow light curves (upper right panel of Figure~\ref{fig:CaII_Ratio_Plot_Inc2013}). There is a preference for SNe~Ia with blueshifted \NaiD\ to have broad light curves than the rest of the sample but as discussed in Section \ref{Sec:HostGalaxyDistribution}, this may be driven by the preference for SNe~Ia with blueshifted \NaiD\ features for late-type host galaxies, which produce more SNe~Ia with broad light curves. Similarly to \cite{childress_2013_HighvelocityFeaturesType}, we do not observe a correlation between $\CaIIRatio$ ratio and \textit{B-V} colour at maximum (bottom left panel of Figure~\ref{fig:CaII_Ratio_Plot_Inc2013}). \cite{pan_2014_HostGalaxiesType} identified that SNe~Ia with high $\CaIIRatio$ ratios are preferentially found in low-mass galaxies. Our sample size is small so we can not draw any strong conclusion from our data on this connection (bottom right panel of Figure~\ref{fig:CaII_Ratio_Plot_Inc2013}).

\section{Discussion}
\label{Sec:Discussion}
In this paper, we have presented a new sample of 15 SNe~Ia with medium-resolution X-shooter spectra, where the properties of narrow \NaiD\ features were studied. We have investigated the connection between the presence of narrow \NaiD\ features, broad SN spectral features (in particular high-velocity features), light curve properties, and the host galaxy stellar mass in a full sample of 47 SNe~Ia (including literature events). Now, we further discuss the implications of these results for our understanding of SNe~Ia and their progenitors.

\subsection{Confirmation of an excess of blueshifted \NaiD\ in late-type hosts}
\label{sec:discuss_excess}

The statistical analysis of \NaiD\ absorption shows a strong preference for displaying the feature blueshifted, with an excess of 23 per cent in the new sample and an overall 24 per cent excess in the full sample. These values are consistent with previous results of a lower limit of $\sim$20 per cent in spiral hosts showing an excess \citep{sternberg_2011_CircumstellarMaterialType,maguire_2013_StatisticalAnalysisCircumstellar}. As was also seen in these previous studies, the SNe~Ia displaying these blueshifted features are strongly concentrated in late-type host galaxies (Figs.~\ref{fig:Combined_Ia_HostGalaxy_Distribution_Plot} and \ref{fig:HostMass_vs_NaIDBlue}). Late-type galaxies in general contain more gas and dust than early-type galaxies \citep[e.g.,][]{galliano_2018_InterstellarDustProperties}, thus it is expected that more SNe occurring within late-type galaxies would have spectra showing $\NaiD$ absorption features (both blue and redshifted relative to the zero velocity position of the SN). As such, a population of SNe~Ia occurring within late-type galaxies would be expected to show more narrow absorption features than a population of SNe located in early-type galaxies, even if the properties of the underlying SNe were obtained from the same distribution. However, there is no reason for such a population to have a preference for blueshifted over redshifted $\NaiD$ features.

We have found that the small sample of SNe~Ia hosted by irregular or dwarf galaxies (four events) showed no \NaiD\ absorption features. Irregular galaxies tend to have young stellar populations and display star formation rates comparable to, though with a greater variance, than late-type galaxies, as well as having higher dust contents \citep{hunter_1997_StarFormationIrregular,hunter_2006_MidInfraredImagesStars}. While the sample is small, the apparent difference in the \NaiD\ absorption behaviour of these objects warrants additional study to determine if it is significant and its origin. One difference between these populations would be the lower metallicities of irregular/dwarf galaxies compared to spiral galaxies \citep{grebel_2004_EvolutionaryHistoryLocal}, although why this would result in different \NaiD\ strengths is unclear.

We have confirmed the previous result of \cite{foley_2012_LINKINGTYPEIa} and \cite{maguire_2012_HubbleSpaceTelescope} that SNe~Ia with broader light curves are also more likely to display \NaiD\ than those with narrower light curves (Figure~\ref{fig:Ia_Stretch_Distribution_Plot}). There is a well-known trend that late-type galaxies preferentially host brighter (broader light curve) SNe~Ia \citep[e.g.,][]{sullivan_2010_DependenceTypeIa} so the result that both the galaxy type and the SN light-curve width correlate with the presence of blueshifted \NaiD\ is unsurprising. However, it is unclear what is the driving force behind this correlation. Is it that brighter SNe~Ia with broader light curves that occur in late-time hosts intrinsically show preferentially blueshifted \NaiD\ absorption or is this an age or metallicity effect? Do SNe~Ia in spiral galaxies have a different progenitor population to those in early-type host galaxies?

\subsection{Lack of a link between blueshifted \NaiD\ and high-velocity \CaII\ features}
\label{sec:discuss_naid_caii}

In the previous section, we discussed how SNe~Ia with blueshifted \NaiD\ have broader light curves but this may be driven by their preference for late-type host galaxies and their differing environments. One of the main aims of this study was to investigate, for the first time, if there is a connection between the presence of broad high-velocity \CaII\ features in SNe~Ia and the presence of narrow blueshifted \NaiD\ features. As previously discussed, high-velocity \CaII\ features are ubiquitous in SNe~Ia and we confirm this here finding that the vast majority of the sample require a high-velocity \CaII\ component. We also identify high-velocity \SiII\ features in 12 SNe~Ia in our sample, although these features are not clearly `detached' from the photospheric component. The reason that high-velocity features are interesting to investigate in connection with blueshifted \NaiD\ features is that there are suggestions that the high-velocity \CaII\ features may be at least partially due to ejecta-CSM interaction \citep[e.g.,][]{mazzali_2005_HighVelocityFeaturesUbiquitous,tanaka_2006_ThreeDimensionalModels}. Therefore, if a link between these quantities was identified, it would provide evidence that blueshifted \NaiD\ features are due to CSM also rather than contamination. We remind the reader that the two probes do explore different distances from the SN, with \CaII\ originating at significantly shorter distances than \NaiD. However, when we examine the strength of the high-velocity \CaII\ components \cite[parameterised through the $\CaIIRatio$ of][]{childress_2013_HighvelocityFeaturesType} compared to the blueshifted $\NaiDTwo$ pEQW (\ref{fig:CaII_Ratio_Plot_Inc2013}), we do not identify any clear trend between them. In particular, we also find that a number of SNe~Ia with no \NaiD\ absorption at all have very high $\CaIIRatio$ values, which is difficult to interpret in the context of a common CSM origin for both strong blueshifted \NaiD\ features and strong high-velocity \CaII\ components, as with this interpretation of their origin a strong signature would be expected in both probes rather than singularly. We also find no correlation between the pEQW of the high-velocity \CaII\ features and the pEQW of the blue-shifted \NaiD\ features. 

Since we do not identify a link between the \NaiD\ features and the strength of high-velocity \CaII\ features, this suggests that either one (or both) of these measurements are unrelated to CSM. Alternatively, these features may explore different regions of CSM, with \NaiD\ absorption produced at much larger distances (10$^{16}$ -- 10$^{17}$ cm) from the SN \citep[e.g.,][]{patat_2007_DetectionCircumstellarMaterial} than the CSM potentially probed through strong high-velocity \CaII\ absorption at < 10$^{15}$ cm \citep{gerardy_2004_SN2003duSignatures, mulligan_2018_CompactCircumstellarShell}. An alternative scenario that has been suggested for the high-velocity \CaII\ components is that they are due to a density or abundance enhancement in the SN ejecta \citep{branch_2006_ComparativeDirectAnalysis, blondin_2012_SPECTROSCOPICDIVERSITYTYPE} and thus have no connection to the presence of CSM. 

\subsection{Can the excess of blueshifted \NaiD\ be explained by CSM?}
\label{sec:CSM_vs_ISM}

We have identified a clear excess of blueshifted over redshifted \NaiD\ absorption features in both our new sample (33 per cent) and our full sample (24 per cent). Previous studies that have investigated these features have suggested that the most probable scenario to produce them is from CSM in the immediate vicinity of the SN \citep[e.g.,][]{sternberg_2011_CircumstellarMaterialType,maguire_2012_HubbleSpaceTelescope,phillips_2013_SourceDustExtinction}. This conclusion was based on the original result of \cite{patat_2007_DetectionCircumstellarMaterial}, where time-varying \NaiD\ features were observed in SN~2006X and were suggested to be due to changes in the ionisation field of the CSM caused by the SN radiation. This material was estimated to be located at a radius of $\sim 10^{16}$ cm from the SN. Other SNe~Ia were subsequently identified with these time-varying \NaiD\ features and were also suggested to be due to CSM, most likely related to explosion scenarios involving non-degenerate companion stars \citep{simon_2009_VariableSodiumAbsorption, blondin_2009_SecondCaseVariable, stritzinger_2010_DistanceNGC1316, sternberg_2014_MultiepochHighspectralresolutionObservations}. There are also models of double-degenerate systems that could eject material several thousand years prior to the explosion that interacts with the ISM, and also could potentially produce similar \NaiD\ profiles \citep[e.g.,][]{raskin_2013_TIDALTAILEJECTION,shen_2013_CircumstellarAbsorptionDouble}. However, it is likely that significant fine-tuning of double-degenerate models would be required to produce such material ejections in a large percentage of SNe~Ia, as well as the host galaxy dependence of SNe~Ia showing blueshifted \NaiD\ absorption features towards late-time host galaxies being less easy to explain with this scenario. Double degenerate scenarios not having a strong preference for a particular stellar population compared to single degenerate models that require the younger stellar populations observed in late type galaxies.

In this study, we have investigated a number of SN~Ia properties directly to see if the presence of blueshifted \NaiD\ features are intrinsically linked to SNe~Ia explosions. As discussed in Sections \ref{sec:discuss_excess} and \ref{sec:discuss_naid_caii}, there is a preference for the presence of \NaiD\ in late-type host galaxies, which also results in a correlation with light-curve stretch, although the driving force behind these correlations is not well understood. We also observe the presence of the previously identified relation that SNe~Ia with redder \textit{B-V} colours at maximum have have stronger \NaiD\ features, though we observe this trend only in objects showing purely blueshifted absorption, with objects showing both blue and redshifted absorption found to have consistent \textit{B-V} colours regardless of the strength of their blueshifted \NaiD\ component. However, \cite{phillips_2013_SourceDustExtinction} showed that this is most likely driven by gas in the ISM and not due to CSM.

Related to the above, \cite{phillips_2013_SourceDustExtinction} investigated the connection between the \NaiD\ column densities seen in SNe~Ia and extinction measured from the SN colours, finding that those with stronger \NaiD\ column densities, compared to the standard Milky Way-derived extinction relation, have preferentially blueshifted \NaiD\ features and hence may have an origin in the CSM. However, there was not a clear correlation with some SNe~Ia with identified time-varying \NaiD\ not following the expected relation, suggesting that the situation is more complicated than previously thought and that strong blueshifted \NaiD\ features may be produced in some way other than through CSM (under the assumption that time-varying \NaiD\ are produced by CSM interaction).

The low R$_V$ values seen in some SNe~Ia have been attributed to multiple scatterings in CSM at typical radii (10$^{16}$ -- 10$^{19}$ cm) \citep{wang_2005_DustTypeIa, patat_2006_ReflectionsReflexionsII, goobar_2008_LowCircumstellarDust, amanullah_2011_PERTURBATIONSSNeIa}, while some more recent studies have investigated the presence of blueshifted \NaiD\ features in the context of an origin in the ISM at 10$^{16}$ - 10$^{20}$ cm \citep[e.g][]{bulla_2018_EstimatingDustDistances,bulla_2018_SheddingLightType}.

These ISM-origin models are based on cloud-cloud collisions produced by SN radiation interacting with clouds in the ISM in the vicinity of the SN at $\sim$~10$^{18}$~cm \citep{hoang_2019_RotationalDisruptionDust}. This mechanism would reduce the dust grain sizes (that would explain the lower R$_v$) and could accelerate the clouds towards the observer (that would explain the preference for blueshifted \NaiD\ features). This scenario, however, requires the IS cloud to be relatively close to the SN ($\lesssim$ 1~pc, see Figure~6 of \citealt{giang_2020_TimevaryingExtinctionPolarization}) in order to explain the low Rv values and blueshifted \NaiD\ features observed already at early phases (pre-maximum). An acceleration of the blueshifted material with time would also be expected though has not been observed in SNe Ia, with blueshifted \NaiD\ features having been observed to evolve in intensity but not in velocity. Newer models (\citealt{hoang_2021_VariationDustProperties}) suggest that interaction with SN radiation is not needed at all and these small dust grains could be produced by the ISM radiation alone before the SN explosion, which would resolve the timescale issue above. There are, however, some remaining issues with this model, including the lack of observations for similar polarisation laws in the Milky Way as observed for SNe Ia \citep{cikota_2018_SpectropolarimetryGalacticStars} and a lack of quantitative estimates of the velocities expected for the gas (and not the dust grains) within the ISM that would give rise to the blueshifted \NaiD\ features. These newer ISM only models also fail to explain why we see the clear excess in blueshifted over redshifted \NaiD\ features that is once again seen to be statistically significant in both our new and full samples of now 47 SNe~Ia with the requisite intermediate-resolution spectroscopy around maximum light.

\begin{figure}
    \centering
    \includegraphics[width=8cm]{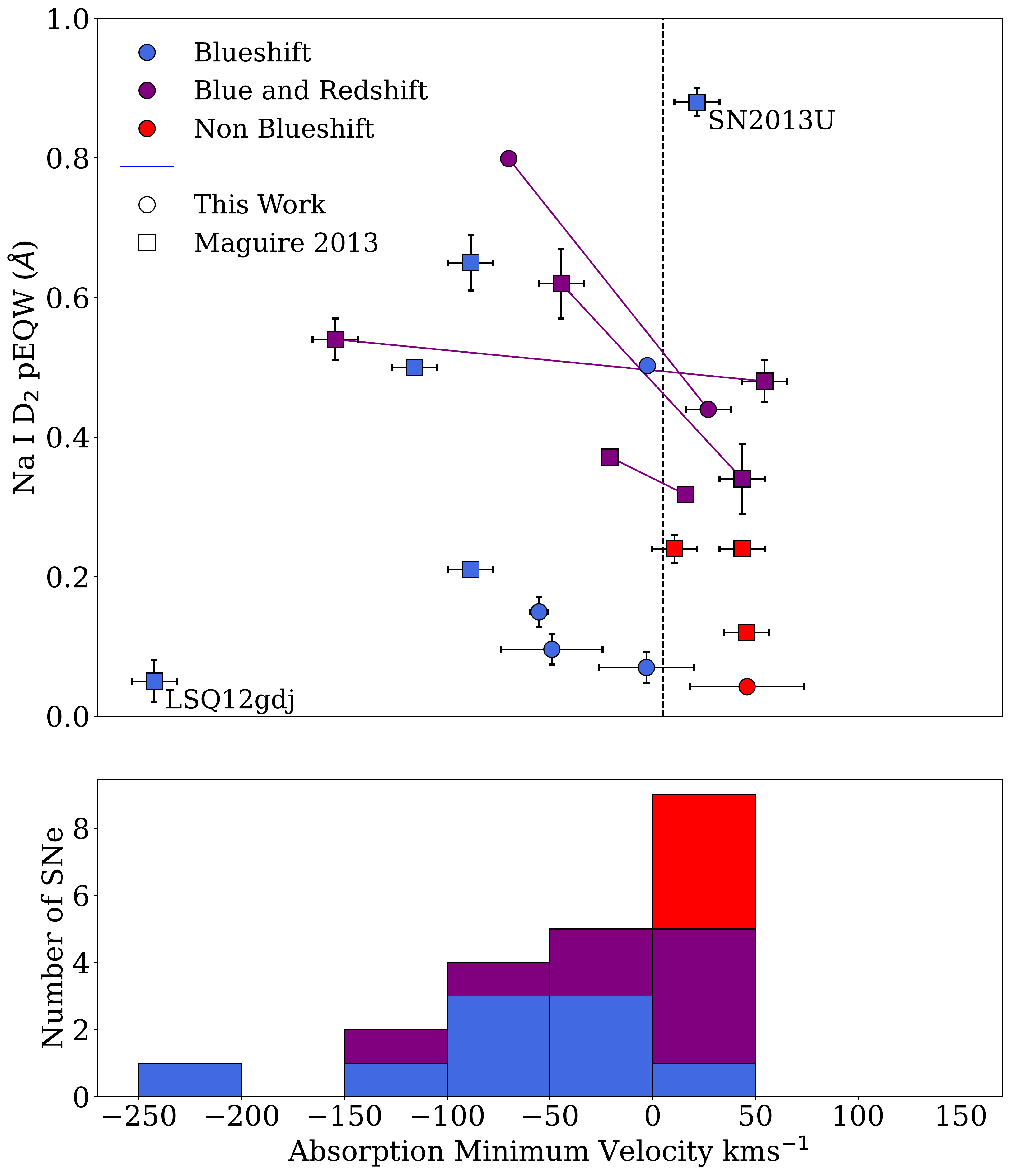}
    \caption{Top panel: \NaiD\ pEQW absorption strength against the absorption minima velocity of the strongest component feature. In the case of `blue and redshifted' objects, pEQW and the absorption minima velocities is measured in two regions (-200 -- 0 \kms\ and 0 -- 200 \kms) joined by connecting lines for the same object. SNe with other absorption classifications are measured across the full feature. Those objects showing no absorption features have been excluded. Objects originally presented in \protect\cite{sternberg_2011_CircumstellarMaterialType} are not included due to differences in the velocity calibration methodology. Bottom panel: The same data presented as a histogram to highlight the differences in distribution between \NaiD\ absorption classification.}
    \label{fig:NaID_pEQW_vs_NaID_Absorption_Minima_Velocity}
\end{figure}

Additionally, there appears to be tentative evidence of a possible distinction in the absorption minimum velocities of the \NaiD\ features between the sub-classifications of absorption feature. With the minima of purely blueshifted absorption features, and the blueshifted components of those objects showing both blue and redshifted absorption, measured to be at higher relative velocities (Mean value: -51$\pm$62~$\kms$) compared to the absorption minima of redshifted absorption features as shown in Figure~\ref{fig:NaID_pEQW_vs_NaID_Absorption_Minima_Velocity} (Mean value: 22$\pm$12~$\kms$). The small number of objects makes drawing firm conclusions based on this difficult but could potentially be explained by a combination of a symmetric population around 0~\kms\ produced by distant ISM with an additional population at higher, bluer relative velocities produced by CSM or close-by ISM.

\section{Conclusions}
\label{Sec:Conclusions}

This work has explored a new sample of moderate-resolution spectra of 15 SNe~Ia obtained near maximum light to investigate the potential link between the properties of narrow \NaiD\ absorption features and high-velocity \CaII\ NIR triplet absorption features, with the goal of constraining the progenitor scenarios of SNe~Ia more firmly. Our main conclusions are:

\begin{enumerate}
    \item we confirmed a \%EB of 23 per cent blueshifted over redshifted \NaiD\ features in a new sample of maximum-light SN~Ia intermediate-resolution spectra, in agreement with previous studies,
    \item we found that this excess was even stronger (48 per cent) in a full sample of 47 SNe~Ia (including literature objects) when only spiral/late-type host galaxies were considered, 
    \item we did not identify a trend between the presence or strength of high-velocity \CaII\ features seen in the majority of SNe~Ia around maximum light and the presence of narrow blueshifted \NaiD\ features,
    \item we observed that the previous identified correlation that stronger blueshifted \NaiD\ features are seen in SNe~Ia displaying redder \textit{B-V} colours at maximum light is present only for those objects displaying purely blueshifted absorption. No trend is observed for objects with blue and redshifted \NaiD\ absorption,
    \item we have identified a tentative relation that the velocity distribution of the narrow blueshifted \NaiD\ features may extend to higher velocities than that of the non-blueshifted \NaiD\ sample
    \item we found that SNe~Ia in irregular galaxies do not show \NaiD\ absorption at odds with their likely underlying stellar populations (although the sample size is very small)
\end{enumerate}

The lack of a link between the properties of the narrow \NaiD\ features and the high-velocity \CaII\ components around maximum light in SNe~Ia suggests either that one or both of these probes is not due to CSM as has previously been suggested or the presence of CSM around Type Ia SNe cannot be linked across the range of distances explored by the two probes.
However, the confirmation of a clear excess of blueshifted over redshifted \NaiD\ absorption features in our new sample, and the full combined literature sample, cannot easily be explained through ISM interpretations of the \NaiD\ feature that do not involve an impact of the SN progenitor system or ejecta on these features. Further theoretical modelling is required to provide new scenarios to explain these results, most likely either in the context of close-in ISM or CSM.

\section*{Acknowledgements}

PC and KM acknowledge support from the EU H2020 ERC grant no. 758638. MB acknowledges support from the Swedish Research Council (Reg. no. 2020-03330). L.G. acknowledges financial support from the Spanish Ministry of Science, Innovation and Universities (MICIU) under the 2019 Ram\'on y Cajal program RYC2019-027683 and from the Spanish MICIU project PID2020-115253GA-I00. SJS acknowledges funding from STFC Grants  ST/P000312/1 and ST/N002520/1.

Based on data taken at the European Organization for Astronomical Research in the Southern Hemisphere, Chile, under program IDs: 098.D-0384(A) and 099.D-0641(A). This research has made use of the NASA/IPAC Extragalactic Database (NED) which is operated by the Jet Propulsion Laboratory, California Institute of Technology, under contract with the National Aeronautics and Space Administration.

This work has made use of data from the Asteroid Terrestrial-impact Last Alert System (ATLAS) project. ATLAS is primarily funded to search for near earth asteroids through NASA grants NN12AR55G, 80NSSC18K0284, and 80NSSC18K1575; byproducts of the NEO search include images and catalogs from the survey area. The ATLAS science products have been made possible through the contributions of the University of Hawaii Institute for Astronomy, the Queen's University Belfast, the Space Telescope Science Institute, the South African Astronomical Observatory (SAAO), and the Millennium Institute of Astrophysics (MAS), Chile.

Funding for the Sloan Digital Sky Survey IV has been provided by the Alfred P. Sloan Foundation, the U.S. Department of Energy Office of Science, and the Participating Institutions. SDSS acknowledges support and resources from the Center for High-Performance Computing at the University of Utah. The SDSS web site is www.sdss.org.

We thank the anonymous reviewer for their helpful comments and discussion.

\section*{Data availability}
The data underlying this article will be shared on reasonable request to the corresponding author.




\bibliographystyle{mnras}
\bibliography{Ia_Progenitor_Probes.bib}


\onecolumn
\appendix

\section{Literature Objects: Broad Feature Fitting Results}
\label{Appendix_Blurb}

For objects identified as not displaying a high velocity feature in their spectrum the `Fitting Notes' column of the following tables details the first criterion each spectrum failed to meet. These criteria are described in more detail in Section~\ref{highvel_needed} but are outlined here for clarity:

\begin{enumerate}
    \item The fitting of a high velocity feature must be successful in at least 10\% of iterations.
    \item The inclusion of a high velocity feature must be strongly favoured by the BIC indicated by a $\Delta\bar{\textrm{BIC}}$ value $>$10. 
\end{enumerate}

\FloatBarrier

\begin{table*}
\caption{Summarised fitting results for the \SiII~6355~\AA\ feature for the objects first described in \protect\cite{maguire_2013_StatisticalAnalysisCircumstellar}. Means have been weighted using the relevant fitting uncertainties with the uncertainties of these means obtained as the standard deviation of the results. A positive value of $\Delta\bar{\textrm{BIC}}$ indicates a preference for the inclusion of a high velocity feature based on statistical considerations alone.}
\label{tab:SiII_Fitting_Results_2013Objects}
\centering
\begin{adjustbox}{max width = \linewidth}
\begin{tabular}{lccccccccc}
\hline
\multicolumn{9}{c}{No High Velocity component} \\ \hline
 &  \multicolumn{3}{c}{Photospheric Component} &  \multicolumn{3}{c}{High Velocity Component} &  &  &\\ \hline
SN	&		Min V (\kms)		&		FWHM (\kms)		&		pEQW (\AA)		&		Min V (\kms)		&		FWHM (\kms)		&		pEQW (\AA)		&		$\bar{\chi}^2$		&		$\Delta\bar{\textrm{BIC}}$& Fitting Notes\\				
LSQ12fxd	&	-11150	$\pm$	50	&	6830	$\pm$	100	&	74.7	$\pm$	1	&		-		&		-		&		-		&	0.251	$\pm$	0.017	&		*		& (1), (i)	\\
PTF12iiq	&	-18980	$\pm$	220	&	10640	$\pm$	880	&	58.4	$\pm$	5.7	&		-		&		-		&		-		&	2.973	$\pm$	0.626	&	-10	$\pm$	140	& (1), (ii)	\\
SN~2012hd	&	-10070	$\pm$	180	&	7510	$\pm$	180	&	31.0	$\pm$	0.0	&		-		&		-		&		-		&	0.119	$\pm$	0.001	&		90 $\pm$ 30		& (1), (i)	\\							
SN~2013U	&	-9980	$\pm$	50	&	7910	$\pm$	70	&	56.3	$\pm$	0.5	&		-		&		-		&		-		&	0.527	$\pm$	0.005	&	-40	$\pm$	40	& (ii)	\\
SN~2013aj	&	-11060	$\pm$	50	&	7750	$\pm$	50	&	118.1	$\pm$	0.1	&		-		&		-		&		-		&	0.113	$\pm$	0.005	&		*		& (i)	\\
\\\hline 												
Mean	&	-11250	$\pm$	3400	&	7650	$\pm$	1310	&	55.6	$\pm$	28.8	&		-		&		-		&		-		&	0.192	$\pm$	1.098	&	0	$\pm$	40	\\ \hline
\\\hline							

\multicolumn{9}{c}{With High Velocity component} \\ \hline
 &  \multicolumn{3}{c}{Photospheric Component} &  \multicolumn{3}{c}{High Velocity Component} &  &  &\\ \hline
SN	&		Min V (\kms)		&		FWHM (\kms)		&		pEQW (\AA)		&		Min V (\kms)		&		FWHM (\kms)		&		pEQW (\AA)		&		$\bar{\chi}^2$		&		$\Delta\bar{\textrm{BIC}}$	& Fitting Notes	\\										LSQ12dbr	&	-11120	$\pm$	400	&	6140	$\pm$	150	&	64.9	$\pm$	8.5	&	-13320	$\pm$	140	&	3020	$\pm$	480	&	10.5	$\pm$	8.5	&	0.130	$\pm$	0.019	&	290	$\pm$	70	&	\\		
LSQ12fuk	&	-9690	$\pm$	50	&	6180	$\pm$	60	&	83.3	$\pm$	0.1	&	-12610	$\pm$	50	&	3070	$\pm$	70	&	8.6	$\pm$	0.1	&	0.074	$\pm$	0.001	&	260	$\pm$	50	&	\\
LSQ12gdj	&	-10910	$\pm$	280	&	5420	$\pm$	250	&	10.3	$\pm$	0.8	&	-12590	$\pm$	210	&	2340	$\pm$	390	&	1.2	$\pm$	0.8	&	0.006	$\pm$	0.001	&	60	$\pm$	20	& (1)	\\
LSQ12hzj	&	-8230	$\pm$	380	&	6890	$\pm$	160	&	51.7	$\pm$	5.4	&	-10440	$\pm$	110	&	4320	$\pm$	360	&	25.1	$\pm$	5.4	&	0.145	$\pm$	0.012	&	340	$\pm$	50	&	\\
PTF12jgb	&	-8950	$\pm$	400	&	5410	$\pm$	370	&	43.4	$\pm$	8.2	&	-11900	$\pm$	200	&	4120	$\pm$	490	&	23.2	$\pm$	8.2	&	0.037	$\pm$	0.002	&	230	$\pm$	80	& (1)	\\
SN~2012cg	&	-9710	$\pm$	120	&	6710	$\pm$	70	&	50.9	$\pm$	3.1	&	-11650	$\pm$	120	&	4660	$\pm$	220	&	24.2	$\pm$	3.1	&	0.041	$\pm$	0.001	&	410	$\pm$	50	&	\\
SN~2012fw	&	-10280	$\pm$	430	&	6030	$\pm$	300	&	11.1	$\pm$	1.9	&	-12160	$\pm$	360	&	3510	$\pm$	700	&	2.6	$\pm$	1.9	&	0.015	$\pm$	0.001	&	50	$\pm$	20	& (1)	\\
SN~2012hr	&	-12470	$\pm$	90	&	8390	$\pm$	100	&	74.1	$\pm$	0.6	&	-16350	$\pm$	200	&	3270	$\pm$	370	&	5.7	$\pm$	0.6	&	0.114	$\pm$	0.025	&	260	$\pm$	40	&	(1)\\
SN~2012ht	&	-9940	$\pm$	60	&	6660	$\pm$	70	&	76.8	$\pm$	1.3	&	-13310	$\pm$	50	&	4890	$\pm$	80	&	33.8	$\pm$	1.3	&	0.045	$\pm$	0.002	&	600	$\pm$	50	& 	\\
SN~2013aa	&	-9750	$\pm$	360	&	5650	$\pm$	250	&	77.4	$\pm$	10.5	&	-12220	$\pm$	190	&	2780	$\pm$	640	&	16.2	$\pm$	10.5	&	0.106	$\pm$	0.014	&	480	$\pm$	100	&	\\							
\\\hline 												
Mean	&	-10250	$\pm$	1120	&	6570	$\pm$	840	&	69	$\pm$	25.1	&	-12650	$\pm$	1460	&	3810	$\pm$	810	&	9.7	$\pm$	10.5	&	0.034	$\pm$	0.047	&	220	$\pm$	160	&
\\\hline																																								
\end{tabular}
\end{adjustbox}
\begin{flushleft}
* No $\Delta\bar{\textrm{BIC}}$ is available for these objects as all fitting iterations when including a high velocity feature were considered to be unsuccessful \\
(1) Values measured from a lower resolution, non X-shooter spectrum \\
Roman numerals denote the first fitting criterion failed by each object not identified as displaying a high velocity feature, as described in Section~\ref{highvel_needed}
\end{flushleft}
\end{table*}

\begin{table*}
\caption{Summarised fitting results for the \CaII\ NIR triplet feature for the objects first described in \protect\cite{maguire_2013_StatisticalAnalysisCircumstellar}. Means have been weighted using the relevant fitting uncertainties with the uncertainties of these means obtained as the standard deviation of the results. A positive value of $\Delta\bar{\textrm{BIC}}$ indicates a preference for the inclusion of a high velocity feature based on statistical considerations alone.}
\label{tab:CaII_Fitting_Results_2013Objects}
\centering
\begin{adjustbox}{max width = \linewidth}
\begin{tabular}{lccccccccc}
\hline
\multicolumn{9}{c}{No High Velocity component} \\ \hline
 &  \multicolumn{3}{c}{Photospheric Component} &  \multicolumn{3}{c}{High Velocity Component} &  &  &\\ \hline
SN	&		Min V (\kms)		&		FWHM (\kms)		&		pEQW (\AA)		&		Min V (\kms)		&		FWHM (\kms)		&		pEQW (\AA)		&		$\bar{\chi}^2$		&		$\Delta\bar{\textrm{BIC}}$ & Fitting Notes\\	\hline			
LSQ12gdj	&	-11000	$\pm$	130	&	3180	$\pm$	140	&	6.0	$\pm$	0.1	&		-		&		-		&		-		&	0.245	$\pm$	0.001	&	40	$\pm$	20	& (1), (i) \\								
PTF12iiq	&	-32170	$\pm$	120	&	32160	$\pm$	310	&	220.5	$\pm$	1.1	&		-		&		-		&		-		&	24.642	$\pm$	0.486	&		*		&(1), (i)\\								
SN~2013U	&	-14440	$\pm$	50	&	16200	$\pm$	60	&	48.4	$\pm$	0.4	&		-		&		-		&		-		&	1.768	$\pm$	0.003	&		*		&(i)\\								
\\\hline 												
Mean	&	-18070	$\pm$	9270	&	14610	$\pm$	11850	&	23.0	$\pm$	92.7	&		-		&		-		&		-		&	0.729	$\pm$	11.159	&		-		&\\ \hline						

\multicolumn{9}{c}{With High Velocity component} \\ \hline
 &  \multicolumn{3}{c}{Photospheric Component} &  \multicolumn{3}{c}{High Velocity Component} &  &  &\\ \hline
SN	&		Min V (\kms)		&		FWHM (\kms)		&		pEQW (\AA)		&		Min V (\kms)		&		FWHM (\kms)		&		pEQW (\AA)		&		$\bar{\chi}^2$		&		$\Delta\bar{\textrm{BIC}}$&	Fitting Notes	\\	\hline														
LSQ12dbr	&	-11040	$\pm$	110	&	4430	$\pm$	40	&	55.3	$\pm$	0.3	&	-18680	$\pm$	40	&	4780	$\pm$	60	&	32.6	$\pm$	0.3	&	1.009	$\pm$	0.007	&	500	$\pm$	60	&   \\
LSQ12fuk	&	-9500	$\pm$	40	&	5300	$\pm$	40	&	104.6	$\pm$	0.1	&	-16840	$\pm$	40	&	4680	$\pm$	40	&	55.3	$\pm$	0.1	&	0.636	$\pm$	0.001	&	1090	$\pm$	60	&	\\
LSQ12fxd	&	-10800	$\pm$	140	&	2810	$\pm$	140	&	9.3	$\pm$	0.2	&	-21340	$\pm$	130	&	5670	$\pm$	140	&	23.2	$\pm$	0.1	&	0.375	$\pm$	0.009	&	270	$\pm$	20	&	(1)\\
LSQ12hzj	&	-8390	$\pm$	300	&	4500	$\pm$	40	&	46.1	$\pm$	0.2	&	-18710	$\pm$	40	&	6040	$\pm$	50	&	29.5	$\pm$	0.2	&	1.354	$\pm$	0.003	&	480	$\pm$	50	&	\\
PTF12jgb	&	-9430	$\pm$	170	&	5280	$\pm$	160	&	67.2	$\pm$	9.6	&	-18510	$\pm$	750	&	5580	$\pm$	1630	&	56.8	$\pm$	9.8	&	0.516	$\pm$	0.245	&	1060	$\pm$	170	& (1)	\\
SN~2012cg	&	-9980	$\pm$	40	&	4440	$\pm$	40	&	51.1	$\pm$	0.1	&	-19510	$\pm$	100	&	5550	$\pm$	40	&	52.5	$\pm$	0.1	&	0.558	$\pm$	0.002	&	1000	$\pm$	60	&	\\
SN~2012fw	&	-9390	$\pm$	190	&	4610	$\pm$	190	&	13.5	$\pm$	0.1	&	-18480	$\pm$	210	&	5590	$\pm$	190	&	11.2	$\pm$	0.1	&	0.406	$\pm$	0.012	&	100	$\pm$	20	&	(1)\\
SN~2012hd	&	-9730	$\pm$	140	&	7390	$\pm$	140	&	47.0	$\pm$	0.1	&	-17490	$\pm$	170	&	5920	$\pm$	140	&	31.6	$\pm$	0.1	&	0.365	$\pm$	0.005	&	240	$\pm$	30	&	(1)\\
SN~2012hr	&	-12400	$\pm$	80	&	8890	$\pm$	140	&	107.2	$\pm$	1.6	&	-20250	$\pm$	120	&	6540	$\pm$	100	&	68.7	$\pm$	1.3	&	1.071	$\pm$	0.026	&	260	$\pm$	50	&	(1)\\
SN~2012ht	&	-10860	$\pm$	40	&	7490	$\pm$	40	&	145.6	$\pm$	0.7	&	-17400	$\pm$	10	&	6960	$\pm$	60	&	67.5	$\pm$	0.7	&	0.629	$\pm$	0.005	&	390	$\pm$	70	&	\\
SN~2013aa	&	-9880	$\pm$	30	&	5700	$\pm$	30	&	119.7	$\pm$	0.1	&	-15840	$\pm$	100	&	4840	$\pm$	40	&	42.9	$\pm$	0.1	&	1.132	$\pm$	0.002	&	340	$\pm$	60	&	\\
SN~2013aj	&	-9810	$\pm$	40	&	6440	$\pm$	40	&	100.1	$\pm$	0.2	&	-17110	$\pm$	40	&	9240	$\pm$	40	&	102	$\pm$	0.2	&	1.288	$\pm$	0.004	&	570	$\pm$	100	&	\\
\\\hline 												
Mean	&	-10180	$\pm$	990	&	5600	$\pm$	1610	&	52.7	$\pm$	41.0	&	-17810	$\pm$	1470	&	6020	$\pm$	1190	&	36.3	$\pm$	23.7	&	0.813	$\pm$	0.352	&	380	$\pm$  320 &
\\\hline
\end{tabular}
\end{adjustbox}
\begin{flushleft}
* No $\Delta\bar{\textrm{BIC}}$ is available as all fitting iterations when including a high velocity feature were considered to be unsuccessful \\
** Residual pattern indicative of the presence of an additional feature \\
(1) Values measured from a lower resolution, non X-shooter spectrum \\
Roman numerals denote the first fitting criterion failed by each object not identified as displaying a high velocity feature, as described in Section~\ref{highvel_needed}
\end{flushleft}
\end{table*}

\begin{table*}
\caption{Summarised fitting results for the \SiII~6355~\AA\ feature for the objects first described in \protect\cite{sternberg_2011_CircumstellarMaterialType}. Means have been weighted using the relevant fitting uncertainties with the uncertainties of these means obtained as the standard deviation of the results. A positive value of $\Delta\bar{\textrm{BIC}}$ indicates a preference for the inclusion of a high velocity feature based on statistical considerations alone.}
\label{tab:SiII_Fitting_Results_SternbergObjects}
\centering
\begin{adjustbox}{max width = \linewidth}
\begin{tabular}{lccccccccc}
\hline
\multicolumn{9}{c}{No High Velocity component} \\ \hline
 &  \multicolumn{3}{c}{Photospheric Component} &  \multicolumn{3}{c}{High Velocity Component} &  &  &\\ \hline
SN	&		Min V (\kms)		&		FWHM (\kms)		&		pEQW (\AA)		&		Min V (\kms)		&		FWHM (\kms)		&		pEQW (\AA)		&		$\bar{\chi}^2$		&		$\Delta\bar{\textrm{BIC}}$& Fitting Notes\\ \hline				
SN~2006cm	&	-11350	$\pm$	90	&	4850	$\pm$	170	&	24.9	$\pm$	0.8	&		-		&		-		&		-		&	0.325	$\pm$	0.012	&		*		&	(1), (i)\\
SN~2008hv	&	-11080	$\pm$	100	&	7450	$\pm$	100	&	48.8	$\pm$	0.3	&		-		&		-		&		-		&	0.037	$\pm$	0.002	&	90	$\pm$	20	&	(1), (i)\\
SN~2008ia	&	-11270	$\pm$	100	&	7140	$\pm$	120	&	53.7	$\pm$	0.6	&		-		&		-		&		-		&	0.080	$\pm$	0.005	&	-60	$\pm$	40	&	(1), (ii)\\
SN~2009ig	&	-13700	$\pm$	210	&	6540	$\pm$	210	&	18.2	$\pm$	0.0	&		-		&		-		&		-		&	0.050	$\pm$	0.001	&		*		&	(1), (i)\\
\\\hline 												
Mean	&	-11560	$\pm$	1070	&	6680	$\pm$	1000	&	22.8	$\pm$	15.1	&		-		&		-		&		-		&	0.061	$\pm$	0.117	&	-10	$\pm$	50	\\ \hline

\multicolumn{9}{c}{With High Velocity component} \\ \hline
 &  \multicolumn{3}{c}{Photospheric Component} &  \multicolumn{3}{c}{High Velocity Component} &  &  &\\ \hline
SN	&		Min V (\kms)		&		FWHM (\kms)		&		pEQW (\AA)		&		Min V (\kms)		&		FWHM (\kms)		&		pEQW (\AA)		&		$\bar{\chi}^2$		&		$\Delta\bar{\textrm{BIC}}$	& Fitting Notes	\\ \hline														SN~2006X	&	-15710	$\pm$	410	&	8420	$\pm$	430	&	134.4	$\pm$	13.9	&	-20430	$\pm$	290	&	4830	$\pm$	540	&	36.2	$\pm$	13.9	&	0.148	$\pm$	0.085	&	1000	$\pm$	170	&	\\
SN~2007on	&	-9930	$\pm$	140	&	6950	$\pm$	150	&	27.3	$\pm$	0.1	&	-13430	$\pm$	140	&	4580	$\pm$	140	&	11.8	$\pm$	0.1	&	0.036	$\pm$	0.001	&	250	$\pm$	20	&	(1)\\
SN~2008fp	&	-10740	$\pm$	150	&	6420	$\pm$	150	&	21.8	$\pm$	0.3	&	-12410	$\pm$	180	&	3460	$\pm$	380	&	2.7	$\pm$	0.3	&	0.023	$\pm$	0.003	&	90	$\pm$	20	&	(1)\\										
\\\hline 												
Mean	&	-11130	$\pm$	2550	&	6940	$\pm$	850	&	26.2	$\pm$	51.9	&	-14590	$\pm$	3570	&	4360	$\pm$	600	&	9.1	$\pm$	14.2	&	0.033	$\pm$	0.056	&	220	$\pm$	400	\\ \hline																																						
\end{tabular}
\end{adjustbox}
\begin{flushleft}
* No $\Delta\bar{\textrm{BIC}}$ is available as all fitting iterations when including a high velocity feature were considered to be unsuccessful \\
(1) Values measured from a lower resolution, non X-shooter spectrum \\
Roman numerals denote the first fitting criterion failed by each object not identified as displaying a high velocity feature, as described in Section~\ref{highvel_needed}
\end{flushleft}
\end{table*}

\begin{table*}
\caption{Summarised fitting results for the \CaII\ NIR triplet feature for the objects first described in \protect\cite{sternberg_2011_CircumstellarMaterialType}. Means have been weighted using the relevant fitting uncertainties with the uncertainties of these means obtained as the standard deviation of the results. A positive value of $\Delta\bar{\textrm{BIC}}$ indicates a preference for the inclusion of a high velocity feature based on statistical considerations alone.}
\label{tab:CaII_Fitting_Results_SternbergObjects}
\centering
\begin{adjustbox}{max width = \linewidth}
\begin{tabular}{lccccccccc}
\hline
\multicolumn{9}{c}{No High Velocity component} \\ \hline
 &  \multicolumn{3}{c}{Photospheric Component} &  \multicolumn{3}{c}{High Velocity Component} &  &  &\\ \hline
SN	&		Min V (\kms)		&		FWHM (\kms)		&		pEQW (\AA)		&		Min V (\kms)		&		FWHM (\kms)		&		pEQW (\AA)		&		$\bar{\chi}^2$		&		$\Delta\bar{\textrm{BIC}}$ & Fitting Notes\\	\hline			
SN~2006cm	&	-11790	$\pm$	70	&	7660	$\pm$	70	&	61.1	$\pm$	0.1	&		-		&		-		&		-		&	1.534	$\pm$	0.007	&	*	& (1), (i)	\\
\\\hline 												
Mean	& - &	-	&	-	&		-		&		-		&		-		&	-	&	-
\\\hline							

\multicolumn{9}{c}{With High Velocity component} \\ \hline
 &  \multicolumn{3}{c}{Photospheric Component} &  \multicolumn{3}{c}{High Velocity Component} &  &  &\\ \hline
SN	&		Min V (\kms)		&		FWHM (\kms)		&		pEQW (\AA)		&		Min V (\kms)		&		FWHM (\kms)		&		pEQW (\AA)		&		$\bar{\chi}^2$		&		$\Delta\bar{\textrm{BIC}}$&	Fitting Notes	\\	\hline													SN~2006X	&	-13330	$\pm$	40	&	6490	$\pm$	50	&	128.1	$\pm$	0.9	&	-20030	$\pm$	40	&	6440	$\pm$	60	&	177.5	$\pm$	1.3	&	0.753	$\pm$	0.027	&	1030	$\pm$	60	&	\\
SN~2007on	&	-8640	$\pm$	160	&	6680	$\pm$	150	&	24.6	$\pm$	2.1	&	-14220	$\pm$	200	&	9360	$\pm$	200	&	51.4	$\pm$	2.1	&	0.143	$\pm$	0.003	&	120	$\pm$	30	&	(1)\\
SN~2008fp	&	-10510	$\pm$	110	&	5320	$\pm$	110	&	20.6	$\pm$	0	&	-17860	$\pm$	150	&	4660	$\pm$	110	&	7.9	$\pm$	0.0	&	0.111	$\pm$	0.004	&	210	$\pm$	20	&	(1)\\
SN~2008hv	&	-9850	$\pm$	70	&	5850	$\pm$	70	&	34.1	$\pm$	0.1	&	-19140	$\pm$	70	&	7050	$\pm$	90	&	20.0	$\pm$	0.1	&	0.241	$\pm$	0.002	&	480	$\pm$	30	&	(1)\\
SN~2008ia	&	-9970	$\pm$	70	&	5760	$\pm$	70	&	48.1	$\pm$	0.1	&	-17720	$\pm$	120	&	6090	$\pm$	70	&	21.9	$\pm$	0.1	&	0.855	$\pm$	0.003	&	200	$\pm$	30	&	(1)\\
SN~2009ig	&	-13500	$\pm$	160	&	6060	$\pm$	160	&	21.5	$\pm$	0	&	-21650	$\pm$	160	&	4890	$\pm$	160	&	4.5	$\pm$	0.0	&	0.045	$\pm$	0.001	&	240	$\pm$	10	&	(1)\\												
\\\hline 												
Mean	&	-11330	$\pm$	1820	&	6040	$\pm$	450	&	28	$\pm$	37.9	&	-19070	$\pm$	2310	&	6290	$\pm$	1560	&	11.5	$\pm$	60.2	&	0.233	$\pm$	0.322	&	290	$\pm$	310&	\\ \hline
\end{tabular}
\end{adjustbox}
\begin{flushleft}
* No $\Delta\bar{\textrm{BIC}}$ is available as all fitting iterations when including a high velocity feature were considered to be unsuccessful \\
** Residual pattern indicative of the presence of an additional feature \\
(1) Values measured from a lower resolution, X-shooter spectrum \\
Roman numerals denote the first fitting criterion failed by each object not identified as displaying a high velocity feature, as described in Section~\ref{highvel_needed}
\end{flushleft}
\end{table*}



\bsp	
\label{lastpage}
\end{document}